\documentclass[11pt,onecolumn,draft]{IEEEtran}

\usepackage{epsf,psfrag,amssymb,amsfonts,cite,amsmath,eufrak,yfonts}
\newtheorem{theorem}{Theorem}
\newtheorem{example}{Example}

\newtheorem{lemma}{Lemma}
\newtheorem{definition}{Definition}

\newtheorem{proposition}{Proposition}

\def\psfancypar#1#2{\begingroup\def\par{\endgraf\endgroup\lineskiplimit=0pt}
               \setbox2=\hbox{\large\sc #2}
               \newdimen\tmpht \tmpht \ht2 \advance\tmpht by \baselineskip
               \font\hhuge=Times-Bold at \tmpht
               \setbox1=\hbox{{\hhuge #1}}
               \count7=\tmpht \count8=\ht1
               \divide\count8 by 1000 \divide\count7 by \count8
               \tmpht=.001\tmpht\multiply\tmpht by \count7
               \font\hhuge=Times-Bold at \tmpht
               \setbox1=\hbox{{\hhuge #1}}
               \noindent
                \hangindent1.05\wd1
               \hangafter=-2 {\hskip-\hangindent
               \lower1\ht1\hbox{\raise1.0\ht2\copy1}%
                \kern-0\wd1}\copy2\lineskiplimit=-1000pt}

\newenvironment{ap}{
\renewcommand{\thesection}{\mbox{Appendix }\Alph{section}}
\renewcommand{\thesubsection}{\Alph{section}.\arabic{subsection
}}
\renewcommand{\theequation}{\mbox{\Alph{section}.}\arabic{equation}}
}{
\renewcommand{\thesection}{\arabic{section}}
\renewcommand{\thesubsection}{\thesection.\arabic{subsection}}
\renewcommand{\theequation}{\arabic{equation}}
}

\newcommand{\beq}{\begin{equation}}
\newcommand{\eeq}{\end{equation}}
\newcommand{\bqa}{\begin{eqnarray}}
\newcommand{\eqa}{\end{eqnarray}}
\newcommand{\bqn}{\begin{eqnarray*}}
\newcommand{\eqn}{\end{eqnarray*}}
\newcommand{\nn}{\nonumber}

\newcommand{\be}{\begin{enumerate}}
\newcommand{\ee}{\end{enumerate}}
\newcommand{\bi}{\begin{itemize}}
\newcommand{\ei}{\end{itemize}}
\newcommand{\bd}{\begin{description}}
\newcommand{\ed}{\end{description}}
\newcommand{\ba}{\begin{array}}
\newcommand{\ea}{\end{array}}
\newcommand{\bde}{\begin{definition}}
\newcommand{\ede}{\end{definition}}
\newcommand{\bex}{\begin{example}}
\newcommand{\eex}{\end{example}}

\newcommand{\Phibf}{\mbox{${\bf \Phi}$}}

\def\boxit#1{\vbox{\hrule\hbox{\vrule\kern3pt
        \vbox{\kern3pt#1\kern3pt}\kern3pt\vrule}\hrule}}

\def\reals{ { {\rm  I \kern-0.15em R }  } }
\def\complex{ {\,{{\rm C} \kern-0.50em \raise0.20ex {  |}}\, }}
\def\alphabf{\hbox{\boldmath$\alpha$\unboldmath}}

\def\gammabf{\hbox{\boldmath$\gamma$\unboldmath}}

\def\Sigmabf{\hbox{$\bf \Sigma$}}

\def\Lambdabf{\mbox{$ \bf \Lambda $}}
\def\Sigmabf{\mbox{$ \bf \Sigma $}}

\def\0bf{{\bf 0}}
\def\1bf{{\bf 1}}
\def\2bf{{\bf 2}}
\def\3bf{{\bf 3}}
\def\4bf{{\bf 4}}
\def\5bf{{\bf 5}}
\def\6bf{{\bf 6}}
\def\7bf{{\bf 7}}
\def\8bf{{\bf 8}}
\def\9bf{{\bf 9}}

\def\Abf{{\bf A}}
\def\Bbf{{\bf B}}
\def\Cbf{{\bf C}}

\def\Ebf{{\bf E}}

\def\Gbf{{\bf G}}
\def\Hbf{{\bf H}}
\def\Ibf{{\bf I}}

\def\Mbf{{\bf M}}

\def\Pbf{{\bf P}}
\def\Qbf{{\bf Q}}
\def\Rbf{{\bf R}}
\def\Sbf{{\bf S}}

\def\Ubf{{\bf U}}
\def\Vbf{{\bf V}}
\def\Wbf{{\bf W}}
\def\Xbf{{\bf X}}

\def\np{{\pmb n}}
\def\up{{\pmb u}}
\def\vp{{\pmb v}}
\def\wp{{\pmb w}}
\def\xp{{\pmb x}}
\def\yp{{\pmb y}}
\def\zp{{\pmb z}}

\def\Bmat{\mathcal{B}}
\def\Cmat{\mathcal{C}}

\def\Nmat{\mathcal{N}}

\def\Pmat{\mathcal{P}}

\def\Smat{\mathcal{S}}
\def\Tmat{\mathcal{T}}

\def\QED{\mbox{\rule[0pt]{1.3ex}{1.3ex}}}

\def\Rxx{\Rbf_{\ssstyle X\kern-.1em X}}

\let\ssstyle=\scriptscriptstyle

\def\etal{{\it et al. \/}}

\def\Cov{{\textrm{Cov}}}
\def\tr{{\textrm{tr}}}
\def\rank{{\textrm{rank}}}
\def\diag{{\textrm{diag}}}

\def\Kout{\setbox1=\hbox{\Huge\bf K}\hbox to
1.05\wd1{\hspace{.05\wd1}
}}
\def\Sout{\setbox1=\hbox{\Huge\bf S}\hbox to 1.05\wd1{\hspace{.05\wd1}
}}

\def\scalefig#1{\epsfxsize #1\textwidth}
 \allowdisplaybreaks[4]

\begin{document}
\title{Capacity Regions and Sum-Rate Capacities of Vector
Gaussian Interference Channels}
\author{Xiaohu Shang, Biao Chen, Gerhard Kramer and H. Vincent Poor\thanks{X.
Shang was with Syracuse University. He is now with Princeton University, Department of Electrical Engineering, Princeton,
NJ, 08544. Email: xshang@princeton.edu. B.
Chen is with Syracuse University, Department of Electrical
Engineering and Computer Science, 335 Link Hall, Syracuse, NY
13244. Email: bichen@syr.edu. G. Kramer was with Bell Labs,
Alcatel-Lucent. He is now with University of Southern California,
Department of Electrical Engineering, 3740 McClintock Ave, Los
Angeles, CA 90089. Email: gkramer@usc.edu.  H. V. Poor is with Princeton University,
Department of Electrical Engineering, Princeton,
NJ, 08544. Email: poor@princeton.edu.
This work was presented in part
in the 46th Annual Allerton Conference, Monticello, IL, Sep. 2008.}} \maketitle
{\footnotetext{This work was supported in part by the National
Science Foundation under Grants CNS-06-25637 and CCF-05-46491.}}

\begin{abstract}

The capacity regions of vector, or multiple-input multiple-output, Gaussian
interference channels are established for very strong interference
and aligned strong interference. Furthermore, the sum-rate
capacities are established for Z interference, noisy interference,
and mixed (aligned weak/intermediate and aligned strong) interference. These
results generalize known results for scalar Gaussian interference
channels.

\end{abstract}

\section{Introduction}

The interference channel (IC) models the situation in which
transmitters communicate with their respective receivers while
generating interference to all other receivers. This channel model
was mentioned in \cite[Section 14]{Shannon:61Berkeley} and its
capacity region is still generally unknown.

In \cite{Carleial:75IT} Carleial showed that interference does not
reduce capacity when it is very strong. This result follows
because the interference can be decoded and subtracted at each
receiver before decoding the desired message. Later Han and
Kobayashi \cite{Han&Kobayashi:81IT} and Sato \cite{Sato:81IT}
showed that the capacity region of the strong interference channel
is the same as the capacity region of a compound multiple access
channel. In these cases, the interference is fully decoded at
both receivers.

\begin{figure}[h] \centerline{
\begin{psfrags}
\psfrag{x1}[c]{$\xp_1$} \psfrag{x2}[c]{$\xp_2$}
\psfrag{y1}[c]{$\yp_1$} \psfrag{y2}[c]{$\yp_2$}
\psfrag{n1}[c]{$\zp_1$} \psfrag{n2}[c]{$\zp_2$} \psfrag{+}[c]{$+$}
\psfrag{g11}[c]{$\Hbf_1$} \psfrag{g12}[c]{$\Hbf_3$}
\psfrag{g21}[c]{$\Hbf_2$} \psfrag{g22}[c]{$\Hbf_4$}
\scalefig{.35}\epsfbox{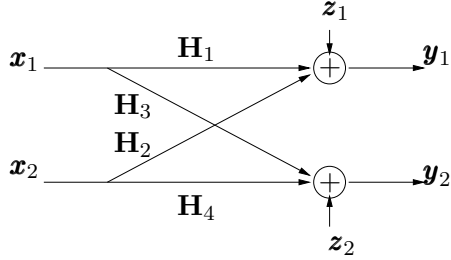}
\end{psfrags}}
\caption{\label{fig:model} The two-user MIMO IC.}
\end{figure}

When the interference is not strong, the capacity region is
unknown. The best inner bound is by Han and Kobayashi
\cite{Han&Kobayashi:81IT}, which was later simplified by Chong
\etal in \cite{Chong-etal:08IT} and \cite{Kramer:06Zurich}. Etkin \etal and
Telatar and Tse showed that Han and Kobayashi's inner bound is
within one bit of the capacity region for scalar Gaussian ICs
\cite{Etkin-etal:08IT} and \cite{Telatar&Tse:07ISIT}. Various outer bounds
have been developed in
\cite{Kramer:04IT,Etkin-etal:08IT,Telatar&Tse:07ISIT,Shang-etal:09IT,
Motahari&Khandani:09IT,Annapureddy&Veeravalli:08IT_submission}.

Special ICs such as the degraded IC and the Z interference channel
(ZIC) were studied in \cite{Sato:78IT} and \cite{Costa:85IT}. Costa proved
that the capacity regions of degraded ICs and ZICs
are the same for the scalar Gaussian
case \cite{Costa:85IT}. The sum-rate capacity for the ZIC was
established  in \cite{Sato:78IT} and \cite{Sason:04IT}.  A recent result in
\cite{Shang-etal:09IT,
Motahari&Khandani:09IT,Annapureddy&Veeravalli:08IT_submission}
showed that if a two-user Gaussian scalar IC has noisy
interference, then treating interference as noise can achieve the
sum-rate capacity. This result has been extended to multi-user
Gaussian ICs in
\cite{shang-etal:08ISIT} and \cite{Annapureddy&Veeravalli:08IT_submission}.
The sum-rate capacity for mixed interference, i.e., one receiver
has strong interference and the other has weak/intermediate interference, was
derived in \cite{Motahari&Khandani:09IT} and
\cite{Weng&Tuninetti:08ITA}.

In this paper, we study the capacity of the two-user Gaussian vector IC or
multiple-input multiple-output (MIMO) IC. As shown in Fig.
\ref{fig:model}, the received signals are defined as
 \bqa
&&\hspace{-.2in}\yp_1=\Hbf_1\xp_1+\Hbf_2\xp_2+\zp_1\quad\textrm{and}\nn\\
&&\hspace{-.2in}\yp_2=\Hbf_3\xp_1+\Hbf_4\xp_4+\zp_2,%
\label{eq:model}%
\eqa
where $\xp_i,i=1,2,$ is the transmitted (column) vector signal of user $i$ which
is subject to the average covariance matrix constraint
\bqa%
\sum_{j=1}^nE\left[\xp_{ij}\xp_{ij}^\dagger\right]\preceq
n\Sbf_i, \label{eq:CovConstraint}
\eqa%
where $\xp_{i1},\xp_{i2},\ldots,\xp_{in}$, is the transmitted vector
sequence of user $i$, and $\Sbf_i$ is a fixed positive
semi-definite matrix. Inequality $\Abf\preceq\Bbf$ means that $\Abf-\Bbf$ is Hermitian positive semi-definite.
The noise $\zp_i$ is a circularly symmetric complex
Gaussian random vector with zero mean and identity covariance
matrix; and $\Hbf_{k}, j=1,\dots,4,$ are the complex channel
matrices known at both the transmitters and receivers. Transmitter $i$ has $t_i$ antennas and receiver $i$ has $r_i$ antennas.

For the MIMO IC, Telatar and Tse \cite{Telatar&Tse:07ISIT} showed
that Han and Kobayashi's region is within one bit per receive
antenna of the capacity region. Some outer bounds for the capacity
region were discussed in \cite{Vishwanath&Jafar:04ITW} and some
lower bounds for the sum-rate capacity based on Han and
Kobayashi's region were given in \cite{Shang-etal:06IT}.
Recent work in \cite{Shang-etal:08Allerton} and \cite{shang:thesis} extended
the existing capacity results from scalar ICs to MIMO ICs under
average power constraints. Specifically,
\cite{Shang-etal:08Allerton} and \cite{shang:thesis} derived the capacity
region for aligned strong interference, and the sum-rate capacity for Z
interference, noisy interference and mixed interference under
average power constraints. It should be noted that some of the results in
\cite{Shang-etal:08Allerton} and \cite{shang:thesis} require the channel
matrices to be square and invertible, and the noisy-interference
sum-rate capacity is obtained by requiring {\em all} possible covariance matrices of $\xp_1$ and $\xp_2$ to satisfy a condition.
 A partially strengthened noisy interference condition for MIMO ICs was later presented in \cite{Annapureddy-etal:08Asi}  which required only that the optimizing covariance matrices of $\xp_1$ and $\xp_2$ satisfy the condition of \cite{Shang-etal:08Allerton} and \cite{shang:thesis}, as long as these
optimizing covariance matrices have full rank (see \cite[Remarks 2 and 3 and Theorem 1]{Annapureddy-etal:08Asi}). A special case
of the MIMO IC, the so-called parallel Gaussian IC where the $\Hbf_i$'s are all square and diagonal
matrices, was studied
in \cite{Shang-etal:08Globecom} and \cite{Shang-etal:09IT_submission}, and it was shown that under suitable conditions for
channel matrices and the power constraints, separate
coding among antennas (or the transmit vector
entries) and treating interference as noise
achieves the sum-rate capacity. In addition, the optimal covariance matrices
 can be singular for this special case. Using the result of \cite{Shang&Chen:07Asi}
that beamforming is optimal for the single-user detection rate
region of the multiple-input single-output (MISO) IC,
\cite{Annapureddy-etal:08Asi} derived noisy-interference
sum-rate capacities for symmetric MISO ICs, i.e., the $\Hbf_j$,
$j=1,\cdots,4$, are all row vectors with $\Hbf_1=\Hbf_4$,
$\Hbf_2=\Hbf_3$ and the two users have identical power
constraints.

 In this paper, we use the covariance matrix constraint
 (\ref{eq:CovConstraint}) and derive the sum-rate capacity of the MIMO IC
with noisy interference, mixed aligned interference,
as well as one-sided interference. The capacity regions of the MIMO IC with
very strong interference and aligned strong interference are also obtained. For all the results,
$\Sbf_i$, $i=1,2$, can be any positive semi-definite matrix,
and the channel matrices $\Hbf_j$, $j=1,\cdots,4$, can be singular
or non-square unless otherwise specified.

The rest of the paper is organized as follows: we present our main
results and numerical examples in Sections II and III, and the proofs of the main results are
given in Section IV.

Before proceeding, we introduce some notation that will be used in
the paper.
\bi %
\item Italic letters (e.g. $X$) denote scalars; and
bold letters $\xp$ and $\Xbf$ denote column vectors and matrices,
respectively.
\item $\Ibf$ denotes the identity matrix
and $\0bf$ denotes the all-zero matrix.
\item $|\Xbf|$, $\Xbf^\dagger$ and $\Xbf^{-1}$ denote
respectively the determinant, conjugate transpose,
 and inverse of the matrix $\Xbf$, and $\|\xp\|$ denotes the Euclidean vector norm of $\xp$.
\item $\textrm{radius}(\Xbf)$ is the numerical radius \cite[p.g. 321]{Horn&Johnson:book} of the square matrix $\Xbf$, and is defined as
\bqn
\textrm{radius}(\Xbf)=\max_{\alphabf^\dagger\alphabf\leq 1}abs\left(\alphabf^\dagger\Xbf\alphabf\right),
\eqn
where $\alphabf$ is a complex vector, and $abs(\cdot)$ denotes the absolute value.
\item
$\xp^n=\left[\xp_1^\dagger,\xp_2^\dagger,\dots,\xp_n^\dagger\right]^\dagger$ is a long
vector which consists of a sequence of vectors $\xp_i, i=1,\dots,
n$.
\item
$\xp\sim\Cmat\Nmat\left(\0bf,\Sigmabf\right)$ means that the random
vector $\xp$ has the circularly symmetric complex Gaussian distribution with zero mean and covariance
matrix $\Sigmabf$.
\item $E[\cdot]$ denotes expectation;
$\textrm{Cov}(\cdot)$ denotes covariance matrix; $I(\cdot;\cdot)$
denotes mutual information; $h(\cdot)$ denotes differential
entropy with the logarithm base $e$, and
$\log(\cdot)=\log_e(\cdot)$.

\ei

\section{Main Results}

In this section, we give the capacity regions for MIMO ICs under
very strong interference and aligned strong interference, and the sum-rate capacities for MIMO ICs
under Z interference, noisy interference and mixed interference.

For economy of notation, we introduce a set of matrices
\bqa%
\Bmat_i=\left\{\Bbf\left|\textrm{all columns of } \Bbf^\dagger
\textrm{ are in the null space of } \Sbf_i\right.\right\},\quad i=1,2,
\eqa
i.e., each column of $\Bbf^\dagger$ is either a zero vector, or an
 eigenvector of the covariance matrix constraint $\Sbf_i$ associated with the zero eigenvalue (if $\Sbf_i$ has one).
 This condition is equivalent to the condition $\Sbf_i\Bbf^\dagger=\0bf$.

\subsection{Capacity region of MIMO IC under very strong interference}

We begin with the result for the MIMO ZIC (MIMO IC with one-sided interference) with very strong interference.
\begin{theorem}
For the MIMO IC defined in (\ref{eq:model}) with $\Hbf_3=0$, if
\bqa%
\log\left|\Ibf+\Hbf_1\Sbf_1\Hbf_1^\dagger+\Hbf_2\Sbf_2\Hbf_2^\dagger\right|-
\log\left|\Ibf+\Hbf_1\Sbf_1\Hbf_1^\dagger\right|&{}\geq{}&\log\left|\Ibf+\Hbf_4\Sbf_2\Hbf_4^\dagger\right|,
\label{eq:Zvstrong}%
\eqa
then the capacity region of the MIMO IC is
\bqa%
\left\{(R_1,R_2):\quad
  0\le R_1\leq\log\left|\Ibf+\Hbf_1\Sbf_1\Hbf_1^\dagger\right|, \quad
  0\le R_2\leq\log\left|\Ibf+\Hbf_4\Sbf_2\Hbf_4^\dagger\right|
\right\},%
\label{eq:Zvestrong} %
\eqa
where $\Sbf_1$ and $\Sbf_2$ are the respective covariance matrix constraints defined in (\ref{eq:CovConstraint}).
\label{theorem:ZICveryStrong}
\end{theorem}

We say that a MIMO ZIC has {\it very strong interference} if
(\ref{eq:Zvstrong}) is satisfied. In this case the interference
does not reduce the capacity region. Theorem \ref{theorem:ZICveryStrong} can be easily extended to obtain the
capacity region for a two-sided MIMO IC under very strong interference.

\begin{theorem}
For the MIMO IC defined in (\ref{eq:model}) and $\Hbf_2\neq\0bf$ and
$\Hbf_3\neq\0bf$, if
\bqa%
\log\left|\Ibf+\Hbf_1\Sbf_1\Hbf_1^\dagger+\Hbf_2\Sbf_2\Hbf_2^\dagger\right|-
\log\left|\Ibf+\Hbf_1\Sbf_1\Hbf_1^\dagger\right|&{}\geq{}&\log\left|\Ibf+\Hbf_4\Sbf_2\Hbf_4^\dagger\right|
\label{eq:vstrong1}\\
\log\left|\Ibf+\Hbf_3\Sbf_1\Hbf_3^\dagger+\Hbf_4\Sbf_2\Hbf_4^\dagger\right|-
\log\left|\Ibf+\Hbf_4\Sbf_2\Hbf_4^\dagger\right|&{}\geq{}&\log\left|\Ibf+\Hbf_1\Sbf_1\Hbf_1^\dagger\right|,%
\label{eq:vstrong2}%
\eqa
then the capacity region of the MIMO IC is
\bqa%
\left\{
  (R_1,R_2): \quad 0\le R_1\leq\log\left|\Ibf+\Hbf_1\Sbf_1\Hbf_1^\dagger\right|, \quad
  0\le R_2\leq\log\left|\Ibf+\Hbf_4\Sbf_2\Hbf_4^\dagger\right|
\right\},%
\label{eq:vestrong} %
\eqa
where $\Sbf_1$ and $\Sbf_2$ are the respective covariance matrix constraints defined in (\ref{eq:CovConstraint}).
\label{theorem:veryStrong}
\end{theorem}

Inequalities (\ref{eq:vstrong1}) and (\ref{eq:vstrong2}) are the
{\em very strong interference} conditions for a two-sided MIMO IC, which
means that when both users transmit at the maximum rate, both
receivers can first decode the interference by treating the desired
signal as noise, i.e., we have
\bqn %
I\left(\xp_2^*;\yp_1^*\right)&{}\geq{}&
I\left(\xp_2^*;\yp_2^*\left|\hspace{.05in}\xp_1^*\right.\right)\quad\textrm{and}\\
I\left(\xp_1^*;\yp_2^*\right)&{}\geq{}&
I\left(\xp_1^*;\yp_1^*\left|\hspace{.05in}\xp_2^*\right.\right),
\eqn
where $\xp_i^*\sim\Cmat\Nmat\left(\0bf,\Sbf_i\right)$ and
$\yp_i^*$ is defined in (\ref{eq:model}) with $\xp_i$ replaced
by $\xp_i^*$, $i=1,2$. As with the scalar Gaussian IC where the notion of very strong interference depends on both the channel coefficients and power constraints, for the MIMO IC our definition of very strong interference involves both the channel matrices and the covariance matrix constraints.  Let $\Hbf_1=\Hbf_4=1$, $\Hbf_2=\sqrt{a}$,
$\Hbf_3=\sqrt{b}$, $\Sbf_1=P_1$ and $\Sbf_2=P_2$, then
(\ref{eq:vstrong1}) and (\ref{eq:vstrong2}) become $a\geq 1+P_1$
and $b\geq 1+P_2$, respectively. Therefore, Theorem
\ref{theorem:veryStrong} generalizes the capacity region for
scalar Gaussian ICs under very strong interference
\cite{Carleial:75IT}.

We remark that an alternative definition of MIMO with very strong interference is to use the power constraints instead of the the covariance matrix constraints. The conditions as well the corresponding capacity region have the same expression as that of Theorem \ref{theorem:veryStrong} except that $\Sbf_1$ and $\Sbf_2$ are now replaced with the waterfilling covariance matrices for the two intended links in the absence of interference. This alternative definition gives a capacity region that is a superset of that defined using the covariance matrix constraints with the trace of the covariance matrices being equal to the power constraints. This alternative definition also includes the scalar Gaussian ICs under very strong interference as its special case.

\subsection{Capacity region of MIMO IC under aligned strong interference}

We begin with the result for the MIMO ZIC under aligned strong interference.
\begin{theorem}
For the MIMO IC defined in (\ref{eq:model}) with $\Hbf_3=0$, if
there exist matrices $\Abf$ and $\Bbf$ such that
\bqa%
\Hbf_4&{}={}&\Abf\Hbf_2+\Bbf,%
\label{eq:ZStrongMarkov}
\eqa
where $\Abf^\dagger\Abf\preceq\Ibf$ and $\Bbf\in\Bmat_2$, then the capacity region of the MIMO IC is
\bqa
 \left\{\begin{array}{c}
  0\leq R_1\leq\log\left|\Ibf+\Hbf_1\Sbf_1\Hbf_1^\dagger\right| \\
  0\leq R_2\leq\log\left|\Ibf+\Hbf_4\Sbf_2\Hbf_4^\dagger\right| \\
  R_1+R_2\leq\log\left|\Ibf+\Hbf_1\Sbf_1\Hbf_1^\dagger+\Hbf_2\Sbf_2\Hbf_2^\dagger\right| \\
\end{array}\right\},%
\label{eq:ZICstrongRegion} %
\eqa
where $\Sbf_1$ and $\Sbf_2$ are the respective covariance matrix constraints defined in (\ref{eq:CovConstraint}).
\label{theorem:ZICstrongRegion}%
\end{theorem}

Theorem \ref{theorem:ZICstrongRegion} gives the capacity region of
a MIMO ZIC under {\em aligned strong interference}. If $\Sbf_2$ is singular, then
(\ref{eq:ZStrongMarkov}) means that all the columns of
$\Hbf_4^\dagger-\Hbf_2^\dagger\Abf^\dagger$ are either zero vectors or the eigenvectors of
$\Sbf_2$ associated with eigenvalue $0$. If $\Sbf_2$ is
nonsingular, then $\Hbf_4=\Abf\Hbf_2$, i.e., $\Hbf_4$ is a linear
transformation of $\Hbf_2$. Therefore, users $1$ and $2$ see
$\xp_2$ in the forms of $\Hbf_2\xp_2$  and $\Abf\Hbf_2\xp_2$,
respectively. If $\Abf^\dagger\Abf\preceq\Ibf$, then
user 1 can decode $\xp_2$ if user 2 can.

The following is a special case of Theorem \ref{theorem:ZICstrongRegion} where we can choose $\Abf$
explicitly as
\bqa
\Abf=\left(\Hbf_4-\Bbf\right)\left(\Hbf_2^\dagger\Hbf_2\right)^{-1}\Hbf_2^\dagger.
\eqa
\begin{proposition}
For the MIMO IC defined in (\ref{eq:model}) with $\Hbf_3=0$,
 if $\Hbf_2$ is left-invertible, i.e., has full column rank, and there exists $\Bbf\in\Bmat_2$ such that
\bqa
\Hbf_2^\dagger\Hbf_2\succeq\left(\Hbf_4-\Bbf\right)^\dagger\left(\Hbf_4-\Bbf\right),%
\label{eq:strongZsimp}
\eqa
then the capacity region of the MIMO IC is given by (\ref{eq:ZICstrongRegion}).%
\label{prop:ZICstrongRegion}%
\end{proposition}

By choosing $\Bbf_i=\0bf$, (\ref{eq:strongZsimp}) becomes $\Hbf_2^\dagger\Hbf_2\succeq\Hbf_4^\dagger\Hbf_4$,
 which is related only to $\Hbf_2$ and $\Hbf_4$ and directly mimics that of the scalar Gaussian IC.

 Using Theorem \ref{theorem:ZICstrongRegion}, we obtain the capacity
 region for the two-sided MIMO IC under aligned strong interference.

\begin{theorem}
For the MIMO IC defined in (\ref{eq:model}), if there exist
matrices $\Abf_1$, $\Abf_2$, $\Bbf_1$ and $\Bbf_2$ such that
\bqa%
\Hbf_1&{}={}&\Abf_1\Hbf_3+\Bbf_1\quad\textrm{and}
\label{eq:strongMarkov1}\\
\Hbf_4&{}={}&\Abf_2\Hbf_2+\Bbf_2,
\label{eq:strongMarkov2}
\eqa
where $\Abf_i^\dagger\Abf_i\preceq\Ibf$ and $\Bbf_i\in\Bmat_i$, $i=1,2$,
then the capacity region of the MIMO IC is
\bqa%
\left\{\begin{array}{c}
  0\leq R_1\leq\log\left|\Ibf+\Hbf_1\Sbf_1\Hbf_1^\dagger\right| \\
  0\leq R_2\leq\log\left|\Ibf+\Hbf_4\Sbf_2\Hbf_4^\dagger\right| \\
  R_1+R_2\leq\log\left|\Ibf+\Hbf_1\Sbf_1\Hbf_1^\dagger+\Hbf_2\Sbf_2\Hbf_2^\dagger\right| \\
  R_1+R_2\leq\log\left|\Ibf+\Hbf_3\Sbf_1\Hbf_3^\dagger+\Hbf_4\Sbf_2\Hbf_4^\dagger\right| \\
\end{array}\right\},%
\label{eq:GICstrongRegion}%
\eqa
where $\Sbf_1$ and $\Sbf_2$ are the respective covariance matrix constraints defined in (\ref{eq:CovConstraint}).
\label{theorem:ICstrongRegion}
\end{theorem}

Similarly to Proposition \ref{prop:ZICstrongRegion}, we have the
following proposition.

\begin{proposition}
For the MIMO IC defined in (\ref{eq:model}), and where the channel
matrices $\Hbf_2$ and $\Hbf_3$ are both left-invertible, if there exist $\Bbf_i\in\Bmat_i$, $i=1,2$, such that
\bqa
\Hbf_2^\dagger\Hbf_2&{}\succeq{}&\left(\Hbf_4-\Bbf_2\right)^\dagger\left(\Hbf_4-\Bbf_2\right)\quad\textrm{and}
\label{eq:strongPorp1}\\
\Hbf_3^\dagger\Hbf_3&{}\succeq{}&\left(\Hbf_1-\Bbf_1\right)^\dagger\left(\Hbf_1-\Bbf_1\right),
\label{eq:strongProp2}
\eqa
 then the capacity region of the MIMO IC is given by
(\ref{eq:GICstrongRegion}).%
\label{prop:ICstrongRegion}%
\end{proposition}

Obviously, Proposition \ref{prop:ICstrongRegion} generalizes the
capacity region of the scalar Gaussian ICs under strong
interference. Furthermore, Proposition \ref{prop:ICstrongRegion}
also generalizes the result of \cite{Vishwanath&Jafar:04ITW} for
single-input multiple-output (SIMO) ICs under strong interference.
In this case, $\Hbf_2$ and $\Hbf_3$ are both non-zero column
vectors, and hence are left-invertible. Therefore, (\ref{eq:strongPorp1}) and (\ref{eq:strongProp2}) become
$\Hbf_2^\dagger\Hbf_2\succeq\Hbf_4^\dagger\Hbf_4$ and
$\Hbf_3^\dagger\Hbf_3\succeq\Hbf_1^\dagger\Hbf_1$ which are the same as
$\|\Hbf_2\|\geq\|\Hbf_4\|$ and $\|\Hbf_3\|\geq\|\Hbf_1\|$.

Let $\Bbf_1=\Bbf_2=\0bf$ and assume that there exist $\Abf_1$ and $\Abf_2$ satisfying (\ref{eq:strongMarkov1}) and (\ref{eq:strongMarkov2}). We can verify
Theorem \ref{theorem:ICstrongRegion} in a way similar to that done in
\cite{Han&Kobayashi:81IT} and \cite{Sato:81IT} for scalar Gaussian ICs under strong interference. Assuming the rate
pair $(R_1,R_2)$ is achievable, then $\xp_1$ and $\xp_2$ can
be reliably recovered at user $1$ and user $2$, respectively.
After subtracting $\xp_1$ from $\yp_1$, user $1$ obtains
\bqa%
\yp_1^\prime=\Hbf_2\xp_2+\zp_1.%
\label{eq:strongYprime}
\eqa
We can pre-multiply $\yp_1^\prime$
by $\Abf_2$ and get
\bqa%
\yp_1^{\prime\prime}&{}={}&\Abf_2\Hbf_2\xp_2+\Abf_2\zp_1\nn\\
&{}={}&\Hbf_4\xp_2+\Abf_2\zp_1.%
\label{eq:strongYdprime}%
\eqa
Since $\xp_1$ is recovered at user $1$, we can add $\Hbf_3\xp_1$
to (\ref{eq:strongYdprime}). Thus user $1$ can eventually compute
\bqa%
\yp_1^{\prime\prime\prime}&{}={}&\Hbf_3\xp_1+\Hbf_4\xp_2+\Abf_2\zp_1.
\label{eq:strongYtprime} %
\eqa
If $\Abf_2^\dagger\Abf_2\preceq\Ibf$, by Lemma \ref{lemma:pd} we have $\Abf_2\Abf_2^\dagger\preceq\Ibf$ and the received signal at user $2$ can be written as
\bqa%
\yp_2&{}={}&\Hbf_3\xp_1+\Hbf_4\xp_2+\zp_2\nn\\
&{}={}&\yp_1^{\prime\prime\prime}+\wp,
\label{eq:strongY2} %
\eqa
where
$\wp\sim\Cmat\Nmat\left(\0bf,\Ibf-\Abf_2\Abf_2^\dagger\right)$,
and $\wp$ is independent of all other random vectors. Since
$\xp_2$ can be recovered from $\yp_2$, $\xp_2$ can also be recovered
from $\yp_1^{\prime\prime\prime}$. Thus, user $1$ can decode both
$\xp_1$ and $\xp_2$. Similarly, user $2$ can also decode both
$\xp_1$ and $\xp_2$. Therefore, the MIMO IC is now a compound MIMO
multiple-access channel, whose capacity region
coincides with (\ref{eq:GICstrongRegion}) \cite{Wyner:74IT}. The above development imposes no
structure on $\xp_i$, $i=1,2$. Therefore, as long as the input signal $\xp_i$ (which can be non-Gaussian
with arbitrary covariance matrix) can be decoded by its desired receiver, it can also be decoded by the other receiver. This result applies to MIMO ICs under a variety of power constraints, for example, peak power constraints, average power constraints and per-antenna power constraints. We state this formally in the following proposition.
\begin{proposition}
For the MIMO IC defined in (\ref{eq:model}) with expected per-symbol power constraints, or expected block power constraints, or per-antenna expected block power constraints, if there exist matrices $\Abf_i$, $i=1,2$, such that $\Abf_i^\dagger\Abf_i\preceq\Ibf$ and
\bqa
&&\Hbf_1=\Abf_1\Hbf_3\quad\textrm{and}\label{eq:strongGeneral1}\\
&&\Hbf_4=\Abf_2\Hbf_2,\label{eq:strongGeneral2}
\eqa
then the capacity region of the MIMO IC is
\bqa
\bigcup_{\left(\widehat\Sbf_1,\widehat\Sbf_2\right)\in\Pmat}\left\{\begin{array}{c}
  0\leq R_1\leq\log\left|\Ibf+\Hbf_1\widehat\Sbf_1\Hbf_1^\dagger\right| \\
  0\leq R_2\leq\log\left|\Ibf+\Hbf_4\widehat\Sbf_2\Hbf_4^\dagger\right| \\
  R_1+R_2\leq\log\left|\Ibf+\Hbf_1\widehat\Sbf_1\Hbf_1^\dagger+\Hbf_2\widehat\Sbf_2\Hbf_2^\dagger\right| \\
  R_1+R_2\leq\log\left|\Ibf+\Hbf_3\widehat\Sbf_1\Hbf_3^\dagger+\Hbf_4\widehat\Sbf_2\Hbf_4^\dagger\right| \\
\end{array}\right\},%
\eqa
where $\Pmat$ denotes the specified power constraints.
\label{prop:generalStrong}
\end{proposition}

For this result, we say that there is an {\it expected per-symbol power} constraint,
an {\it expected block power} constraint, and a {\it per-antenna expected block power} constraint, respectively,
if the following conditions must be satisfied:
\bqa
&&\tr\left(E\left[\xp_{ij}\xp_{ij}^\dagger\right]\right)\leq P_i,\quad j=1,\cdots,n,\\
&&\sum_{j=1}^n\tr\left(E\left[\xp_{ij}\xp_{ij}^\dagger\right]\right)\leq nP_i\quad\textrm{or}\\
&&\sum_{j=1}^n\left(E\left[\xp_{ij}\xp_{ij}^\dagger\right]\right)_{k}\leq nP_{ik},
\eqa
where $(\cdot)_k$ denotes the $k$th diagonal element of a square matrix, and $P_{ik}$ is the power constraint for the $k$th antenna of user $i$.

Theorem \ref{theorem:ICstrongRegion} has relaxed conditions on the channel matrices as compared to Proposition \ref{prop:generalStrong}. The extra term $\Bbf_i$ in Theorem \ref{theorem:ICstrongRegion} results from the covariance matrix constraint $\Sbf_i$. Suppose (\ref{eq:strongMarkov1}) and (\ref{eq:strongMarkov2}) hold and the input signal of user $i$ is $\xp_i^*\sim\Cmat\Nmat\left(\0bf,\Sbf_i\right)$. From Theorem \ref{theorem:ICstrongRegion}, $\xp_i^*$ achieves the capacity. Applying the same procedure in (\ref{eq:strongYprime})-(\ref{eq:strongYtprime}) to $\yp_1$, we obtain the counterpart of (\ref{eq:strongYdprime})
\bqa
\bar\yp^{\prime\prime}&{}={}&\Abf_2\Hbf_2\xp_2^*+\Abf_2\zp_1\nn\\
&{}={}&\left(\Abf_2\Hbf_2+\Bbf_2\right)\xp_2^*+\Abf_2\zp_1\nn\\
&{}={}&\Hbf_4\xp_2^*+\Abf_2\zp_1,
\eqa
where the second equality holds since
\bqa
\Cov\left(\Bbf_2\xp_2^*\right)=\Bbf_2\Sbf_2\Bbf_2^\dagger=\0bf,
\eqa
and hence $\Bbf_2\xp_2^*=\0bf$. Therefore, $\yp_2$ can also be written as (\ref{eq:strongY2}).

The difference between Proposition \ref{prop:generalStrong} and Theorem \ref{theorem:ICstrongRegion} is that (\ref{eq:strongGeneral1}) and (\ref{eq:strongGeneral2}) ensure that $\xp_i$ can be reliably decoded at both receivers as long as it can be decoded at the desired receiver, while (\ref{eq:strongMarkov1}) and (\ref{eq:strongMarkov2}) ensure that the capacity-achieving $\xp_i^*$ can be reliably decoded at both receivers.

\subsection{Sum-rate capacity of MIMO IC under noisy interference}

In \cite{Shang-etal:09IT}, we say that an IC has {\em noisy
interference} when treating interference as noise achieves the
sum-rate capacity. In this section, we present the sum-rate
capacity results for MIMO ICs that have noisy interference.

\begin{theorem}
For the MIMO IC defined in (\ref{eq:model}) with $\Hbf_3=\0bf$, if
there exist matrices $\Abf$ and $\Bbf$ that satisfy
\bqa%
\Hbf_2&{}={}&\Abf^\dagger\Hbf_4+\Bbf,
\label{eq:ZICweakMarkov}
\eqa
where $\Abf^\dagger\Abf\preceq\Ibf$ and $\Bbf\in\Bmat_2$, then the sum-rate capacity of the MIMO IC is
\bqa%
\log\left|\Ibf+\Hbf_1\Sbf_1\Hbf_1^\dagger\left(\Ibf+\Hbf_2\Sbf_2\Hbf_2^\dagger\right)^{-1}\right|
+\log\left|\Ibf+\Hbf_4\Sbf_2\Hbf_4^\dagger\right|,%
\label{eq:ZICweakSum}
\eqa
where $\Sbf_1$ and $\Sbf_2$ are the respective covariance matrix constraints defined in (\ref{eq:CovConstraint}).
\label{theorem:ZGICsum}
\end{theorem}

Similarly to Proposition \ref{prop:ZICstrongRegion}, we obtain the
following result.

\begin{proposition}
For the MIMO IC defined in (\ref{eq:model}) with $\Hbf_3=\0bf$, if
$\Hbf_4$ is left-invertible and there exists $\Bbf\in\Bmat_2$ such that
\bqa
\Hbf_4^\dagger\Hbf_4\succeq\left(\Hbf_2-\Bbf\right)^\dagger\left(\Hbf_2-\Bbf\right),
\label{eq:weakZsimp}
\eqa
 then the sum-rate capacity of the MIMO IC is given by (\ref{eq:ZICweakSum}). %
\label{prop:ZICSum}
\end{proposition}

Theorem \ref{theorem:ZGICsum} gives the noisy-interference
sum-rate capacity of a MIMO ZIC. Specifically, when
(\ref{eq:ZICweakMarkov}) is satisfied,
the sum-rate capacity can be achieved by treating interference as
noise. Consider a scalar Gaussian IC where $\Hbf_1=\Hbf_4=1$,
$\Hbf_2=\sqrt{a}$ and $\Hbf_3=0$. Equation
(\ref{eq:ZICweakMarkov}) is $0\leq a\leq 1$. Therefore, Theorem
\ref{theorem:ZGICsum} includes the scalar Gaussian ZIC
noisy-interference sum-rate capacity as a special case\footnote{The case with $a<1$ is often referred to as ZIC with weak interference in the literature. We use the term noisy-interference simply because of the fact that treating interference as noise achieves the sum-rate capacity.}. For a SIMO
IC where $\Hbf_1$, $\Hbf_3$ and $\Hbf_4$ are all nonzero column
vectors, Proposition \ref{prop:ZICSum} shows that if
$\|\Hbf_2\|\leq \|\Hbf_4\|$, the sum-rate capacity is achieved by
treating interference as noise.

Similarly to Proposition \ref{prop:generalStrong}, if we choose $\Bbf=\0bf$ in (\ref{eq:ZICweakMarkov}), then Theorem \ref{theorem:ZGICsum} can be extended for different power constraints. We state this formally in the following proposition.
\begin{proposition}
For the MIMO IC defined in (\ref{eq:model}) with expected per-symbol power constraints, or expected block power constraints, or per-antenna expected block power constraints, if $\Hbf_3=\0bf$ and there exists a matrix $\Abf$ such that $\Abf^\dagger\Abf\preceq\Ibf$ and
\bqa
\Hbf_2=\Abf^\dagger\Hbf_4,
\label{eq:ZICgeneral}
\eqa
then the sum-rate capacity is
\bqa%
\max_{\left(\widehat\Sbf_1,\widehat\Sbf_2\right)\in\Pmat}
\left(\log\left|\Ibf+\Hbf_1\widehat\Sbf_1\Hbf_1^\dagger\left(\Ibf+\Hbf_2\widehat\Sbf_2\Hbf_2^\dagger\right)^{-1}\right|
+\log\left|\Ibf+\Hbf_4\widehat\Sbf_2\Hbf_4^\dagger\right|\right),%
\label{eq:ZICweakGeneral}
\eqa
where $\Pmat$ denotes the specified power constraints.
\label{prop:generalZICweak}
\end{proposition}

Next, we give the noisy-interference sum-rate capacity of a
two-sided MIMO IC. Note that this result does not require $\Sbf_1$
or $\Sbf_2$ to have full rank (see \cite{Annapureddy-etal:08Asi} and Example \ref{ex:MISONI} below).

\begin{theorem}
For the MIMO IC defined in (\ref{eq:model}),
 if there exist matrices $\Abf_i$, $\Bbf_i\in\Bmat_i$, and Hermitian positive
definite matrices $\Sigmabf_i$, $i=1,2$, such that
\bqa%
&&\Abf_1^\dagger\Abf_1\preceq\Sigmabf_1\preceq\Ibf-\Abf_{2}\Sigmabf_{2}^{-1}\Abf_{2}^\dagger,%
\label{eq:condition2_1}\\
&&\Abf_2^\dagger\Abf_2\preceq\Sigmabf_2\preceq\Ibf-\Abf_1\Sigmabf_1^{-1}\Abf_1^\dagger,%
\label{eq:condition2_2}\\
&&\Hbf_3=\Abf_1^\dagger\left(\Hbf_2\Sbf_2\Hbf_2^\dagger+\Ibf\right)^{-1}\Hbf_1+\Bbf_1\quad\textrm{and}%
\label{eq:NI2A1}\\
&&\Hbf_2=\Abf_2^\dagger\left(\Hbf_3\Sbf_1\Hbf_3^\dagger+\Ibf\right)^{-1}\Hbf_4+\Bbf_2,%
\label{eq:NI2A2}
\eqa
then the sum-rate capacity of the MIMO IC is
\bqa%
\log\left|\Ibf+\Hbf_1\Sbf_1\Hbf_1^\dagger\left(\Ibf+\Hbf_2\Sbf_2\Hbf_2^\dagger\right)^{-1}\right|
+\log\left|\Ibf+\Hbf_4\Sbf_2\Hbf_4^\dagger\left(\Ibf+\Hbf_3\Sbf_1\Hbf_3^\dagger\right)^{-1}\right|,%
\label{eq:sumcapacity}%
\eqa
where $\Sbf_1$ and $\Sbf_2$ are the respective covariance matrix constraints defined in (\ref{eq:CovConstraint}).
\label{theorem:NIsum2}
\end{theorem}

Theorem \ref{theorem:NIsum2} gives sufficient conditions for
the MIMO IC under which treating interference as noise achieves
the sum-rate capacity. In the case where both $\Hbf_1$
and $\Hbf_4$ are left-invertible, the following conditions are
sufficient for (\ref{eq:NI2A1}) and (\ref{eq:NI2A2}):
\bqa%
\Abf_1&{}={}&\left(\Ibf+\Hbf_2\Sbf_2\Hbf_2^\dagger\right)\Hbf_1
\left(\Hbf_1^\dagger\Hbf_1\right)^{-1}\left(\Hbf_3^\dagger-\Bbf_1^\dagger\right)\quad\textrm{and}%
\label{eq:A1simplified}\\
\Abf_2&{}={}&\left(\Ibf+\Hbf_3\Sbf_1\Hbf_3^\dagger\right)\Hbf_4
\left(\Hbf_4^\dagger\Hbf_4\right)^{-1}\left(\Hbf_2^\dagger-\Bbf_2^\dagger\right).%
\label{eq:A2simplified}%
\eqa
That is, such matrices $\Abf_1$ and $\Abf_2$ exist when $\Hbf_1$
and $\Hbf_4$ are left-invertible. It remains to find matrices $\Bbf_1\in\Bmat_1$ and $\Bbf_2\in\Bmat_2$
such the matrix inequalities (\ref{eq:condition2_1}) and (\ref{eq:condition2_2}) have solutions. We state this formally in the following proposition.

\begin{proposition}
For the MIMO IC defined in (\ref{eq:model}), if $\Hbf_1$ and
$\Hbf_4$ are left-invertible, and there exist symmetric positive
definite matrices $\Sigmabf_1$ and $\Sigmabf_2$ that satisfy
(\ref{eq:condition2_1}) and (\ref{eq:condition2_2}) with $\Abf_1$
and $\Abf_2$ defined in (\ref{eq:A1simplified}) and
(\ref{eq:A2simplified}) for some $\Bbf_1\in \Bmat_1$ and $\Bbf_2\in \Bmat_2$, then the
sum-rate capacity is given by (\ref{eq:sumcapacity}). %
\label{prop:NI2}
\end{proposition}

Although Theorem \ref{theorem:NIsum2} gives the noisy interference conditions for a MIMO IC, finding explicit solution of the matrix inequalities (\ref{eq:condition2_1}) and (\ref{eq:condition2_2}) can be very complex. Therefore, using Theorem \ref{theorem:NIsum2} to check whether a MIMO IC has noisy interference is not practical.
We thus derive the following proposition that is a
special case of Theorem \ref{theorem:NIsum2} but is more amenable to computation.

\begin{proposition}
For the MIMO IC defined in (\ref{eq:model}), the sum-rate capacity
is given by (\ref{eq:sumcapacity}) if
\bqa
\textrm{radius}\left(\Phibf_i\right)\leq\frac{1}{2},\quad i=1,2,
\label{eq:condition}
%
\eqa
where
\bqa%
\Phibf_1&{}={}&\left(\Ibf-\Abf_1^\dagger\Abf_1-\Abf_2\Abf_2^\dagger\right)^{-\frac{1}{2}}\Abf_1^\dagger\Abf_2^\dagger
\left(\Ibf-\Abf_1^\dagger\Abf_1-\Abf_2\Abf_2^\dagger\right)^{-\frac{1}{2}}\\
\Phibf_2&{}={}&\left(\Ibf-\Abf_1\Abf_1^\dagger-\Abf_2^\dagger\Abf_2\right)^{-\frac{1}{2}}\Abf_2^\dagger\Abf_1^\dagger
\left(\Ibf-\Abf_1\Abf_1^\dagger-\Abf_2^\dagger\Abf_2\right)^{-\frac{1}{2}},
\eqa
and $\Abf_1$ and $\Abf_2$ are chosen to satisfy
(\ref{eq:NI2A1}) and
(\ref{eq:NI2A2}) respectively,
and $\Bbf_i\in\Bmat_i$, $i=1,2$.
\label{prop:NIsum}
\end{proposition}

  In the scalar case, if we have
$\Hbf_1=\Hbf_4=1$, $\Hbf_2=\sqrt{a}$, $\Hbf_3=\sqrt{b}$,
$\Sbf_1=P_1$ and $\Sbf_2=P_2$, from (\ref{eq:condition}) we directly have
\bqn
\sqrt{a}(1+bP_1)+\sqrt{b}(1+aP_2)\leq 1.%
\eqn
The above condition can also be obtained from Theorem \ref{theorem:NIsum2} after some mathematical manipulations. Therefore Theorem \ref{theorem:NIsum2} and Proposition \ref{prop:NIsum} generalize the
noisy-interference sum-rate capacity of the scalar Gaussian IC
\cite{Shang-etal:09IT,
Motahari&Khandani:09IT,Annapureddy&Veeravalli:08IT_submission} to
the MIMO IC.

Similarly to Proposition \ref{prop:NI2}, we obtain the following
proposition.

\begin{proposition}
For the MIMO IC defined in (\ref{eq:model}), if both $\Hbf_1$ and
$\Hbf_4$ are left-invertible, and there exist matrices $\Bbf_1\in\Bmat_1$ and $\Bbf_2\in\Bmat_2$ such that the $\Abf_1$ and $\Abf_2$ defined
in (\ref{eq:A1simplified}) and (\ref{eq:A2simplified}) satisfy
(\ref{eq:condition}), then the sum-rate capacity is
(\ref{eq:sumcapacity}).%
 \label{prop:NI1}
\end{proposition}

\subsection{Sum-rate capacity of MIMO IC under mixed aligned interference}

\begin{theorem}
For the MIMO IC defined in (\ref{eq:model}), if there exist
matrices $\Abf_1$, $\Abf_2$, $\Bbf_1$ and $\Bbf_2$ that satisfy
\bqa%
\Hbf_1&{}={}&\Abf_1\Hbf_3+\Bbf_1\quad\textrm{and}
\label{eq:mixedMarkov1}\\
\Hbf_2&{}={}&\Abf_2^\dagger\Hbf_4+\Bbf_2,
\label{eq:mixedMarkov2}
%
%
\eqa
where $\Abf_i^\dagger\Abf_i\preceq\Ibf$ and $\Bbf_i\in\Bmat_i$, $i=1,2$,
 then the sum-rate capacity of the MIMO IC is
\bqa%
\min\left\{\begin{array}{c}
\log\left|\Ibf+\Hbf_3\Sbf_1\Hbf_3^\dagger+\Hbf_4\Sbf_2\Hbf_4^\dagger\right| \\
  \log\left|\Ibf+\Hbf_1\Sbf_1\Hbf_1^\dagger\left(\Ibf+\Hbf_2\Sbf_2\Hbf_2^\dagger\right)^{-1}\right|
  +\log\left|\Ibf+\Hbf_4\Sbf_2\Hbf_4^\dagger\right|\\
  \end{array}\right\},%
\label{eq:GICmixed}%
\eqa
where $\Sbf_1$ and $\Sbf_2$ are the respective covariance matrix constraints defined in (\ref{eq:CovConstraint}).
\label{theorem:GICmixed}
\end{theorem}

\begin{proposition}
For the MIMO IC defined in (\ref{eq:model}) where $\Hbf_3$ and
$\Hbf_4$ are left-invertible, if there exist $\Bbf_i\in\Bmat_i$, $i=1,2$, such that
\bqa%
\Hbf_4^\dagger\Hbf_4&{}\succ{}&\left(\Hbf_2-\Bbf_2\right)^\dagger\left(\Hbf_2-\Bbf_2\right)\quad\textrm{and}
\label{eq:mixed1}\\
\Hbf_3^\dagger\Hbf_3&{}\succeq{}&\left(\Hbf_1-\Bbf_1\right)^\dagger\left(\Hbf_1-\Bbf_1\right),
\label{eq:mixed2}
\eqa
 then the sum-rate capacity is given by (\ref{eq:GICmixed}).%
\label{prop:GICmixed}
\end{proposition}

Theorem \ref{theorem:GICmixed} gives the sum-rate capacity of the
MIMO IC under {\em mixed aligned interference}, i.e., one user sees aligned weak/intermediate
interference and the other user sees aligned strong interference. The
sum-rate capacity is achieved by treating interference as noise at
the receiver that sees aligned weak/intermediate interference, and fully decoding the
interference at the receiver that sees aligned strong interference.
Proposition \ref{prop:GICmixed} includes the sum-rate
capacity of scalar Gaussian ICs with mixed interference as a special case.
If we choose $\Bbf_1=\0bf$ and $\Bbf_2=\0bf$, the constraints (\ref{eq:mixed1}) and (\ref{eq:mixed2}) reduce to $\Hbf_4^\dagger\Hbf_4\succ\Hbf_2^\dagger\Hbf_2$ and $\Hbf_3^\dagger\Hbf_3\succeq\Hbf_1^\dagger\Hbf_1$. The MIMO ICs that satisfy these two simplified conditions have mixed interference and this result applies to channels with other power constraints.

Similar to Propositions \ref{prop:generalStrong} and \ref{prop:generalZICweak}, we obtain the sum-rate capacity for MIMO ICs with aligned mixed interference under different power constraints.
\begin{proposition}
For the MIMO IC defined in (\ref{eq:model}) with expected per-symbol power constraints, or expected block power constraints, or per-antenna expected block power constraints, if there exist matrices $\Abf_i$, $i=1,2$, such that $\Abf_i^\dagger\Abf_i\preceq\Ibf$ and
\bqa
\Hbf_1&{}={}&\Abf_1\Hbf_3\quad\textrm{and}
\label{eq:mICgeneral1}\\
\Hbf_2&{}={}&\Abf_2^\dagger\Hbf_4,
\label{eq:mICgeneral2}
\eqa
then the sum-rate capacity is
\bqa%
\max_{\left(\widehat\Sbf_1,\widehat\Sbf_2\right)\in\Pmat}\min\left\{\begin{array}{c}
\log\left|\Ibf+\Hbf_3\widehat\Sbf_1\Hbf_3^\dagger+\Hbf_4\widehat\Sbf_2\Hbf_4^\dagger\right| \\
  \log\left|\Ibf+\Hbf_1\widehat\Sbf_1\Hbf_1^\dagger\left(\Ibf+\Hbf_2\widehat\Sbf_2\Hbf_2^\dagger\right)^{-1}\right|
  +\log\left|\Ibf+\Hbf_4\widehat\Sbf_2\Hbf_4^\dagger\right|\\
  \end{array}\right\},
\eqa
where $\Pmat$ denotes the specified power constraints.
\label{prop:generalMIC}
\end{proposition}

\subsection{Generalizations}

The results in the previous sections are for MIMO ICs whose capacities are achieved by $\xp_i\sim\Cmat\Nmat\left(\0bf,\Sbf_i\right)$, $i=1,2$, where $\Sbf_i$ is the covariance matrix constraint for user $i$ defined in (\ref{eq:CovConstraint}). The methods introduced can also be applied to more general cases in which the capacity is achieved by $\xp_i^\prime\sim\Cmat\Nmat\left(\0bf,\Sbf_i^\prime\right)$ where $\Sbf_i^\prime\preceq\Sbf_i$. For example, consider the following generalization of Theorem \ref{theorem:ZGICsum} that gives the sum-rate capacity of a class of MIMO ZICs under noisy interference. This extension applies to all the corresponding theorems for other kinds of interference.
\begin{theorem}
For the MIMO IC defined in (\ref{eq:model}), if $\Hbf_3=\0bf$ and the optimal $\Abf^*$, $\Sbf_1^*$ and $\Sbf_2^*$ for the following optimization problem
\bqa
\min_{\Abf}\hspace{.02in}\max_{\widehat\Sbf_1,\widehat\Sbf_2}&&\quad C\left(\Abf,\widehat\Sbf_1,\widehat\Sbf_2\right)\nn\\
\textrm{subject to}&&\quad \Abf\Abf^\dagger\preceq\Ibf,\quad\0bf\preceq\widehat\Sbf_1\preceq\Sbf_1,\quad\0bf\preceq\widehat\Sbf_2\preceq\Sbf_2,
\label{eq:bound}
\eqa
satisfy
\bqa
&&\Hbf_2={\Abf^*}^\dagger\Hbf_4+\Bbf,
\label{eq:tightCond}
\eqa
where
\bqa
\hspace{-.2in}C\left(\Abf,\widehat\Sbf_1,\widehat\Sbf_2\right)&{}={}&\log\left|\Hbf_1\widehat\Sbf_1\Hbf_1^\dagger+\Hbf_2\widehat\Sbf_2\Hbf_2^\dagger+\Ibf\right|
-\log\left|\Ibf-\Abf\Abf^\dagger\right|\nn\\
&&+\log\left|\Hbf_4\widehat\Sbf_2\Hbf_4^\dagger+\Ibf-\left(\Hbf_4\widehat\Sbf_2\Hbf_2^\dagger+\Abf\right)
\left(\Hbf_2\widehat\Sbf_2\Hbf_2^\dagger+\Ibf\right)^{-1}\left(\Hbf_2\widehat\Sbf_2\Hbf_4^\dagger+\Abf^\dagger\right)\right|,
\eqa
and
\bqa
\Bbf\in\left\{\tilde\Bbf\left|\textrm{ all columns of }\tilde\Bbf^\dagger\textrm{ are in the null space of }\Sbf_2^*\right.\right\},
\eqa
then the sum-rate capacity for the MIMO IC is
\bqa
\log\left|\Ibf+\Hbf_1\Sbf_1^*\Hbf_1^\dagger\left(\Ibf+\Hbf_2\Sbf_2^*\Hbf_2^\dagger\right)^{-1}\right|
+\log\left|\Ibf+\Hbf_4\Sbf_2^*\Hbf_4^\dagger\right|.
\label{eq:ZICwekSum}
\eqa
\label{theorem:bound}
\end{theorem}

The solution of problem (\ref{eq:bound}) is an upper bound on the sum-rate capacity of this MIMO ZIC. The bound is tight when (\ref{eq:tightCond}) is satisfied. Theorem \ref{theorem:bound} includes Theorem \ref{theorem:ZGICsum} as a special case in which $\Sbf_1$ and $\Sbf_2$ are optimal for problem (\ref{eq:bound}).

\section{Numerical Results}

\begin{example}

Consider a MIMO IC with
\bqn
\Hbf_1=\Hbf_4=\Ibf,\quad
\Hbf_2=\left
[\begin{array}{cc}
 2.0 &\quad 1.5 \\
 0.8 &\quad 1.0
 \end{array}
\right],\quad
\Hbf_3=\left
[\begin{array}{cc}
 1.2 &\quad 2.0 \\
 0 &\quad 0.8
 \end{array}
\right]\hspace{.1in}\textrm{and}\hspace{.1in}
\Sbf_1=\Sbf_2=\Ibf.
\eqn
Conditions (\ref{eq:vstrong1}) and (\ref{eq:vstrong2}) are satisfied. Therefore this MIMO IC has very strong interference and the capacity region is
\bqn%
\left\{(R_1,R_2):\quad 0\le R_1\leq 1.3863,\quad 0\le R_2\leq 1.3863\right\}.
\eqn
However, consider the aligned strong interference conditions (\ref{eq:strongMarkov1}) and (\ref{eq:strongMarkov2}) for this channel. We have $\Abf_1=\Hbf_3^{-1}$, $\Abf_2=\Hbf_2^{-1}$ and $\Bbf_1=\Bbf_2=\0bf$, where $\Abf_1^\dagger\Abf_1\npreceq\Ibf$ and $\Abf_2^\dagger\Abf_2\npreceq\Ibf$. Therefore, the above channel has very strong interference but not aligned strong interference.

\end{example}

\begin{example}
Consider a MIMO IC with
\bqn%
&&\Hbf_1=\left[\begin{array}{cccc}
  1.8 &\quad 0.8 &\quad -0.6 &\quad 1.4 \\
  1.2 &\quad -1.9 &\quad 0.5 &\quad -0.7 \\
\end{array}\right],\quad
\Hbf_2=\left[\begin{array}{cccc}
  0.8 &\quad 1.0 &\quad -0.5 &\quad 0.6 \\
  1.0 &\quad -1.2 &\quad 0.4 &\quad 1.2 \\
\end{array}\right],\\
&&\Hbf_3=\left[\begin{array}{cccc}
  1.0 &\quad 1.0 &\quad 0.5 &\quad 0.5 \\
  0.4 &\quad 0.2 &\quad 1 &\quad 0.6 \\
\end{array}\right],\hspace{.5in}
\Hbf_4=\left[\begin{array}{cccc}
  0.68 &\quad 0.36 &\quad -0.22 &\quad 0.6 \\
  1.04 &\quad -0.66 &\quad 0.17 &\quad 1.14 \\
\end{array}\right],\\
 &&\Sbf_1=\left[\begin{array}{cccc}
  0.9 &\quad 0.4 &\quad 1.0 &\quad 0.1 \\
  0.4 &\quad 0.4 &\quad 0 &\quad -0.4 \\
  1.0 &\quad 0 &\quad 2.0 &\quad  1.0 \\
  0.1 &\quad -0.4& \quad 1.0 &\quad 0.9\\
\end{array}\right]\hspace{.1in}\textrm{and}\hspace{.1in}
\Sbf_2=\Ibf. %
\eqn
Conditions (\ref{eq:strongMarkov1})-(\ref{eq:strongMarkov2}) are both
satisfied by choosing
\bqn%
\Abf_1=\left[\begin{array}{cc}
  0.8 &\quad 0 \\
  0 &\quad 0.5 \\
\end{array}\right],\quad
\Abf_2=\left[\begin{array}{cc}
  0.6 &\quad 0.2 \\
  0.3 &\quad 0.8 \\
\end{array}\right],\quad
\Bbf_1=\left[\begin{array}{cccc}
1&\quad 0&\quad -1&\quad 1\\
1&\quad-2&\quad 0&\quad -1\\
\end{array}\right]\hspace{.1in}\textrm{and}\hspace{.1in}
\Bbf_2=\0bf.
\eqn
By Theorem
\ref{theorem:ICstrongRegion}, this MIMO IC is under aligned strong
interference and the capacity region is
\bqn%
\left\{(R_1,R_2):\quad 0\le R_1\leq 1.6770,\quad 0\le R_2\leq 1.8636,\quad 0\le R_1+R_2\leq 3.2812\right\} %
\eqn
\label{ex:MISOStrong}
\end{example}

\begin{example}
Consider a MIMO ZIC where
\bqn%
&&\Hbf_1=\Ibf,\quad
\Hbf_2=\left[\begin{array}{cccc}
  1.3 &\quad 1.1 &\quad 1.4 \\
  1.5 &\quad -0.5 &\quad 3.0\\
  0.9 &\quad -0.36 &\quad 1.5\\
\end{array}\right],\quad
\Hbf_3=\0bf,\quad
\Hbf_4=\left[\begin{array}{cccc}
  1.0 &\quad 2.0 &\quad 0.5 \\
  1.0 &\quad 1.0 &\quad 2\\
  0.5 &\quad 0.4 &\quad 0.5\\
\end{array}\right],\\
&&\Sbf_1=\Ibf\hspace{.1in}\textrm{and}\hspace{.1in}
\Sbf_2=\left[\begin{array}{cccc}
  1.8 &\quad 1.0 &\quad -0.4 \\
  1.0 &\quad 5.0 &\quad 2.0\\
  -0.4 &\quad 2.0 &\quad 1.2\\
\end{array}\right].%
\eqn

Condition (\ref{eq:ZICweakMarkov}) is
satisfied by choosing
\bqn
\Abf=\left[\begin{array}{ccc}
             0.8 &\quad0 &\quad0 \\
              0&\quad 0.5 & \quad0 \\
             0&\quad0  &\quad 0.6
           \end{array}
\right]\hspace{.1in}\textrm{and}\hspace{.1in}
\Bbf=\left[\begin{array}{ccc}
             0.5 &\quad -0.5 & \quad 1.0 \\
             1.0 &\quad -1.0 &\quad 2.0 \\
             0.6 &\quad -0.6 &\quad 1.2
           \end{array}
\right].
\eqn
 By Theorem
\ref{theorem:ZGICsum} or Proposition \ref{prop:ZICSum}, the above MIMO ZIC is under noisy
interference and the sum-rate capacity $C=5.6622$ is obtained from
(\ref{eq:ZICweakSum}).

\end{example}

\begin{example}
Consider a MISO IC with
\bqn%
\left[\begin{array}{c}
  \Hbf_1 \\
  \Hbf_2 \\
  \Hbf_3 \\
  \Hbf_4 \\
\end{array}\right]&{}={}&\left[\begin{array}{cccc}
  6.0 &\quad 4.0 &\quad 5.0 \\
  0.5 &\quad 0.8 &\quad 1.0 \\
  0.4 &\quad 0.6 &\quad 0.1 \\
  3.0 &\quad -2.0 &\quad 6.0 \\
\end{array}\right],\quad
\Sbf_1=\left[\begin{array}{ccc}
  0.9 &\quad 0.5 &\quad -0.2 \\
  0.5 &\quad 2.5 &\quad 1 \\
  -0.2 &\quad 1 &\quad 0.6 \\
\end{array}\right]\hspace{.1in}\textrm{and}\hspace{.1in}
\Sbf_2=\left[\begin{array}{ccc}
  2.2 &\quad -0.2 &\quad -0.6 \\
  -0.2 &\quad 0.2 &\quad -0.4 \\
  -0.6 &\quad -0.4 &\quad 1.3 \\
\end{array}\right]. %
\eqn
Condition (\ref{eq:condition}) is
 satisfied by choosing
\bqn
\Abf_1=0.1578,\quad\Abf_2=0.2394,\quad\Bbf_1=[-0.2,0.2,-0.4]\hspace{.1in}\textrm{and}\hspace{.1in}\Bbf_2=[0.2,1.0,0.4].
\eqn
 By
Proposition \ref{prop:NIsum}, this MISO IC is under noisy
interference and the sum-rate capacity $C=7.7171$ is achieved by
treating interference as noise. In this case
$\rank\left(\Sbf_1\right)=\rank\left(\Sbf_2\right)=2$.
However, if we use average power constraints
$P_1=\tr\left(\Sbf_1\right)=4.0$ and
$P_2=\tr\left(\Sbf_2\right)=3.7$ instead of the covariance
matrix constraints $\Sbf_1$ and $\Sbf_2$, then using the
optimality of beamforming for single-user detection of MISO ICs
\cite{Shang&Chen:07Asi}, we can achieve a sum rate of $R_1+R_2=9.9162$
by treating interference as noise and
choosing $\Sbf_i=\gammabf_i\gammabf_i^\dagger$, $\rank\left(\Sbf_i\right)=1$, $i=1,2$, where
$\gammabf_1=[1.2133,-0.0181,1.5899]^\dagger$ and
$\gammabf_2=[0.5673,-1.4460,1.1345]^\dagger$. %
\label{ex:MISONI}
\end{example}

\begin{example}
Consider a MIMO IC under average power constraints $P_1=8$ and $P_2=1$ with
\bqn
\Hbf_1&{}={}&\diag[1.0392,1.5937,1.2689],\hspace{.35in}
\Hbf_2=\diag[0.7746,0.2646,0.3162],\\
\Hbf_3&{}={}&\diag[0.3000,0.6083,0.3162]\hspace{.1in}\textrm{and}\hspace{.1in}
\Hbf_4=\diag[1.5330,1.2124,1.3784].
\eqn
Since all the channel matrices are diagonal, this MIMO IC can be considered as a parallel IC. From \cite[Theorem 3]{Shang-etal:09IT_submission}, this MIMO IC is under noisy interference and the sum-rate capacity  $C=6.1066$ can be achieved by independent coding across antennas and treating interference as noise. The optimal input signals are Gaussian with covariance matrices
\bqn%
\bar\Sbf_1\triangleq\diag\left[2.0922,3.3021,2.6057\right]\hspace{.1in}\textrm{and}\hspace{.1in}
\bar\Sbf_2\triangleq\diag\left[0.4472,0,0.5528\right],
\eqn
where $\tr\left(\bar\Sbf_1\right)=P_1$ and $\tr\left(\bar\Sbf_2\right)=P_2$. The input covariance matrix of the second user is singular and the second antenna is inactive.

If the average power constraints $P_1$ and $P_2$ are replaced by covariance constraints:
\bqn
\Sbf_1&{}={}&
\left[\begin{array}{ccc}
     2.0922 &\quad 0.5000 &\quad 1.0000 \\
     0.5000 &\quad 3.3021 &\quad 0 \\
     1.0000 &\quad 0      &\quad 2.6057
                   \end{array}
\right]\hspace{.1in}\textrm{and}\hspace{.1in}
\Sbf_2=
\left[\begin{array}{ccc}
     0.4472 &\quad 0      &\quad 0.1500 \\
     0      &\quad 0      &\quad 0 \\
     0.1500 &\quad 0      &\quad 0.5528
                   \end{array}
\right],
\eqn
where $\tr(\Sbf_1)=P_1$ and $\tr(\Sbf_2)=P_2$ but $\Sbf_1\nsucceq\bar\Sbf_1$ and  $\Sbf_2\nsucceq\bar\Sbf_2$.
Conditions (\ref{eq:A1simplified}) and (\ref{eq:A2simplified}) are satisfied by choosing
\bqn
\Abf_1=
\left[\begin{array}{ccc}
     0.3661 &\quad 0      &\quad 0.0092 \\
     0      &\quad 0.3817      &\quad 0 \\
     0.0106 &\quad 0      &\quad 0.2630
                   \end{array}
\right],\quad
\Abf_2=
\left[\begin{array}{ccc}
     0.6004 &\quad 0.0199      &\quad 0.0218 \\
     0.0461      &\quad 0.4848      &\quad 0 \\
     0.0479 &\quad 0      &\quad 0.2892
                   \end{array}
\right],\hspace{.1in}\textrm{and}\hspace{.1in}
\Bbf_1=\Bbf_2=\0bf.
\eqn
It can be obtained from (\ref{eq:condition}) that $\textrm{radius}(\Phibf_1)=0.4614$ and $\textrm{radius}(\Phibf_2)=0.1822$. Therefore, from Proposition \ref{prop:NIsum} this MIMO IC is under noisy interference and the sum-rate capacity $C=5.9541$ is achieved by treating interference as noise.
\end{example}

{
\section{Proofs of the Main Results}

We first introduce some lemmas which will be used to prove our main results.

\subsection{Preliminaries}

The following lemma is based on the fact that a Gaussian
distribution maximizes conditional entropy under a covariance
matrix constraint \cite{Thomas:87IT}.
\begin{lemma}%
${}$ Let
$\xp_i^n=\left[\xp_{i,1}^\dagger,\dots,\xp_{i,n}^\dagger\right]^\dagger,i=1,\dots,k$,
be $k$ long random vectors each of which consists of $n$ vectors.
Suppose the $\xp_{i,j}$, $i=1,\cdots,k$ all have the same length
$L_j$, $j=1,\cdots,n$. Let
$\yp^{n}=\left[\yp_1^\dagger,\dots,\yp_n^\dagger\right]^\dagger$, where $\yp_j$ has
length $L_j$, be a long
Gaussian random vector with covariance matrix
\bqa
\Cov\left(\yp^{n}\right)=\sum_{i=1}^k\lambda_i\Cov\left(\xp^n_i\right),\label{eq:cvxCov}
\eqa
where $\sum_{i=1}^k\lambda_i=1,\lambda_i\geq 0$ and $\left|\Cov\left(\xp_i^n\right)\right|>0$. Let $\Smat$ be a
subset of $\{1,2,\dots,n\}$ and $\Tmat$ be a subset of $\Smat$'s
complement. Then we have
\bqa%
 \sum_{i=1}^k\lambda_i
h\left(\xp_{i,\Smat}\left|\xp_{i,\Tmat}\right.\right)\leq
h\left(\yp_\Smat\left|\yp_{\Tmat}\right.\right).%
\eqa
\label{lemma:generalconcave}
\end{lemma}
Proof: See Appendix A.

 When the $\xp_k$, $k=1,\cdots,n$ are all Gaussian distributed, Lemma
\ref{lemma:generalconcave} shows that
$h\left(\xp_\Smat\left|\xp_{\Tmat}\right.\right)$ is concave over
the covariance matrices.

Lemma \ref{lemma:conditionaldirect} includes some special cases of Lemma
\ref{lemma:generalconcave}.

\begin{lemma}
Let $\xp^k=\{\xp_1,\cdots,\xp_k\}$ and $\yp^k=\{\yp_1,\cdots,\yp_k\}$ be two sequences of random vectors, and let $\widehat\xp^*$, $\widehat\yp^*$, $\bar\xp^*$ and $\bar\yp^*$ be Gaussian vectors with covariance matrices satisfying
\bqa
\textrm{Cov}\left[\begin{array}{c}
  \widehat\xp^* \\
  \widehat\yp^* \\
\end{array}\right]=\frac{1}{k}\sum_{i=1}^k\textrm{Cov}\left[\begin{array}{c}
  \xp_i \\
  \yp_i \\
\end{array}\right]\preceq\textrm{Cov}\left[\begin{array}{c}
  \bar\xp^* \\
  \bar\yp^* \\
\end{array}\right].%
\label{eq:cndCov} %
\eqa
Then we have
\bqa %
&&h\left(\xp^k\right)\leq k\cdot h\left(\widehat\xp^*\right)\leq k\cdot h\left(\bar\xp^*\right)\quad\textrm{and}%
\label{eq:cndEntropy1}\\
&&h\left(\yp^k\left|\xp^k\right.\right)\leq k\cdot
h\left(\widehat\yp^*\left|\widehat\xp^*\right.\right) \leq k\cdot
h\left(\bar\yp^*\left|\bar\xp^*\right.\right).%
\label{eq:cndEntropy2} %
\eqa
\label{lemma:conditionaldirect}
\end{lemma}

Proof: See Appendix B.

\begin{lemma}
Let $\xp^n=\{\xp_1,\cdots,\xp_n\}$ be a sequence of $n$ random vectors and let $\bar\xp^*$ and $\widehat\xp^*$ be Gaussian
random vectors with covariance matrices
\bqa%
\Cov\left(\widehat\xp^*\right)=\frac{1}{n}\sum_{i=1}^n\Cov\left(\xp_i\right)\preceq\Cov\left(\bar\xp^*\right).%
\eqa
Let
$\zp$ and $\tilde\zp$ be two independent Gaussian random vectors
and $\zp^n$ and $\tilde\zp^n$ be two sequences of random vectors
each independent and identically distributed (i.i.d.) as $\zp$ and
$\tilde\zp$, respectively. We have
\bqa%
h\left(\xp^n+\zp^n\right)-h\left(\xp^n+\zp^n+\tilde\zp^n\right)&{}\leq{}&
nh\left(\widehat\xp^*+\zp\right)-nh\left(\widehat\xp^*+\zp+\tilde\zp\right)\\
&{}\leq{}&nh\left(\bar\xp^*+\zp\right)-nh\left(\bar\xp^*+\zp+\tilde\zp\right).%
\eqa
\label{lemma:opt}%
\end{lemma}
Proof: See Appendix C.

\begin{lemma}\cite[page 107]{Hajek:book}\cite{Bandemer-etal:08CTW}
Let $\xp,\yp$ and $\zp$ be joint Gaussian vectors. If $\Cov(\yp)$ is invertible, then
$\xp\rightarrow\yp\rightarrow\zp$  forms a Markov chain if and only if
\bqa%
\Cov\left(\xp,\zp\right)=\Cov\left(\xp,\yp\right)\Cov\left(\yp\right)^{-1}\Cov\left(\yp,\zp\right).\nn%
\eqa
\label{lemma:markov}
\end{lemma}

Using Lemma \ref{lemma:markov} we obtain the following lemma.

\begin{lemma}
Let $\xp$, $\up$ and $\vp$ be jointly Gaussian vectors, such that
$\xp$ is independent of $\up$ and $\vp$. Denote
$\Cov\left(\xp\right)=\Sbf_x$, $\Cov\left(\up\right)=\Sbf_u$ and
$\Cov\left(\up,\vp\right)=\Sbf_{uv}$. If $\Sbf_u$ is
invertible, then $\xp\rightarrow\Hbf\xp+\up\rightarrow\Gbf\xp+\vp$
forms a Markov chain if and only if
\bqa%
\Sbf_x\Gbf^\dagger=\Sbf_x\Hbf^\dagger\Sbf_u^{-1}\Sbf_{uv}.%
\label{eq:myMarkovChain} %
\eqa
\label{lemma:myMarkovChain}
\end{lemma}

Proof: See Appendix D.

\begin{lemma} $%
\left[\begin{array}{ll}
  \Ibf &\quad \Abf \\
  \Abf^\dagger &\quad \Bbf \\
\end{array}%
\right]\succeq \0bf$ if and only if $\Bbf\succeq\Abf^\dagger\Abf$. If $\Bbf\succ\0bf$, then $\Bbf\succeq\Abf^\dagger\Abf$ if and only if $\Ibf\succeq\Abf\Bbf^{-1}\Abf^\dagger$.
\label{lemma:pd}
\end{lemma}
Proof: See Appendix E.

\begin{lemma}
If $\Bbf$ is left-invertible (or $\Bbf^\dagger\Bbf$ is invertible) and
$\Abf=\Bbf\left(\Bbf^\dagger\Bbf\right)^{-1}\Cbf^\dagger$, then
$\Abf^\dagger\Abf\preceq\Ibf$ or $\Abf\Abf^\dagger\preceq\Ibf$ if and only if
$\Bbf^\dagger\Bbf\succeq\Cbf^\dagger\Cbf$. %
\label{lemma:leftInvPs}
\end{lemma}
Proof: See Appendix F.

\begin{lemma}\cite[Theorem 5.2]{Engwerda-etal:93LA&A}
Suppose $\Wbf$ is nonsingular and $\Mbf$ is positive definite.
Then the matrix equation
\bqa %
\Xbf+\Wbf^\dagger\Xbf^{-1}\Wbf=\Mbf\nn %
\eqa
has a positive definite solution $\Xbf$ if and only if
\bqa
\textrm{radius}\left(\Mbf^{-\frac{1}{2}}\Wbf\Mbf^{-\frac{1}{2}}\right)\leq\frac{1}{2}.\nn
\eqa
\label{lemma:pdsolution}
\end{lemma}

Using Lemma \ref{lemma:pdsolution}, we obtain necessary and sufficient
conditions for a pair of matrix equations to have
positive definite solutions.

\begin{lemma}
Suppose $\Abf_1$ and $\Abf_2$ are fixed, and $\Ibf$ is the identity
matrix, the following matrix equations
\bqa%
\Sigmabf_1&{}={}&\Ibf-\Abf_2\Sigmabf_2^{-1}\Abf_2^\dagger\quad\textrm{and}
\label{eq:eqt1}\\
\Sigmabf_2&{}={}&\Ibf-\Abf_1\Sigmabf_1^{-1}\Abf_1^\dagger,%
\label{eq:eqt2}%
\eqa
have positive definite solutions $\Sigmabf_1\succ\Abf_1^\dagger\Abf_1$
and $\Sigmabf_2\succ\Abf_2^\dagger\Abf_2$ if and only if
\bqa
%
\textrm{radius}\left(\Phibf_i\right)\leq\frac{1}{2},\quad i=1,2,%
\label{eq:cond}
%
\eqa
where
\bqa%
\Phibf_1&{}={}&\left(\Ibf-\Abf_1^\dagger\Abf_1-\Abf_2\Abf_2^\dagger\right)^{-\frac{1}{2}}\Abf_1^\dagger\Abf_2^\dagger
\left(\Ibf-\Abf_1^\dagger\Abf_1-\Abf_2\Abf_2^\dagger\right)^{-\frac{1}{2}}\quad\textrm{and}%
\label{eq:phi1}\\
\Phibf_2&{}={}&\left(\Ibf-\Abf_1\Abf_1^\dagger-\Abf_2^\dagger\Abf_2\right)^{-\frac{1}{2}}\Abf_2^\dagger\Abf_1^\dagger
\left(\Ibf-\Abf_1\Abf_1^\dagger-\Abf_2^\dagger\Abf_2\right)^{-\frac{1}{2}}.%
\label{eq:phi2}
\eqa
\label{lemma:solutions}
\end{lemma}
Proof: See Appendix G.

\subsection{Proof of Theorem \ref{theorem:ZICveryStrong}}

The converse follows by giving receiver $1$ the message not
destined for it and applying
the maximum-entropy theory to show that Gaussian input distributions
are optimal. To prove
achievability, let $\xp_1\sim\Cmat\Nmat\left(\0bf,\Sbf_1\right)$ and
 $\xp_2\sim\Cmat\Nmat\left(\0bf,\Sbf_2\right)$, and let user $1$
 transmit at rate
 $R_1=\log\left|\Ibf+\Hbf_1\Sbf_1\Hbf_1^\dagger\right|$,
 and user $2$ transmit at rate
 $R_2=\log\left|\Ibf+\Hbf_4\Sbf_2\Hbf_4^\dagger\right|$.
 Inequality (\ref{eq:Zvstrong})
 guarantees that user $1$ can first decode $\xp_2$
  by treating $\xp_1$ as noise. After the interference is
 subtracted, user $1$ sees a single-user Gaussian MIMO channel.
 Therefore, the rate region (\ref{eq:Zvestrong}) is achievable.

\subsection{Proof of Theorem \ref{theorem:veryStrong}}
Similarly to the proof of Theorem \ref{theorem:ZICveryStrong}, the converse follows by giving each receiver the message not
destined for it and applying
the maximum-entropy theory to show that Gaussian input distributions
are optimal. To prove the achievability, let
$\xp_1\sim\Cmat\Nmat\left(\0bf,\Sbf_1\right)$ and
 $\xp_2\sim\Cmat\Nmat\left(\0bf,\Sbf_2\right)$, and let user $1$
 transmit at rate
 $R_1=\log\left|\Ibf+\Hbf_1\Sbf_1\Hbf_1^\dagger\right|$,
 and user $2$ transmit at rate
 $R_2=\log\left|\Ibf+\Hbf_4\Sbf_2\Hbf_4^\dagger\right|$.
 Inequalities (\ref{eq:vstrong1}) and (\ref{eq:vstrong2})
 guarantee that each user can first fully decode the interference
 by treating the desired signals as noise. After the interference is
 subtracted, each user sees a single-user Gaussian MIMO channel.
 Therefore, the rate region (\ref{eq:vestrong}) is achievable.

\subsection{Proof of Theorem \ref{theorem:ZICstrongRegion} and Proposition \ref{prop:ZICstrongRegion}}

Suppose the channel is used $n$ times. The transmitted and
received vector sequences are denoted by $\xp_i^n$ and $\yp_i^n$
for user $i$, $i=1,2$, and $\xp_i^n$ satisfies
(\ref{eq:CovConstraint}).

 Since $\Abf^\dagger\Abf\preceq\Ibf$, from Lemma \ref{lemma:pd}, there exists a Gaussian random
vector $\np$ whose joint distribution with $\zp_2$ is
\bqa%
\left[\begin{array}{c}
  \zp_2 \\
  \np \\
\end{array}\right]\sim\Cmat\Nmat\left(\0bf,\left[\begin{array}{cc}
  \Ibf &\quad \Abf \\
  \Abf^\dagger &\quad \Ibf \\
\end{array}\right]\right).%
\label{eq:ZICjoint2}%
\eqa
Moreover, from (\ref{eq:ZStrongMarkov}), $\np$ is of the same dimension as $\zp_1$ hence has the same marginal distribution as $\zp_1$.

Let $\epsilon>0$ and
$\epsilon\rightarrow 0$ as $n\rightarrow+\infty$, From Fano's inequality, any
 achievable rates must satisfy
\bqa%
&&n(R_1+R_2)-n\epsilon\nn\\
&&\leq
I\left(\xp_1^n;\yp_1^n\right)+I\left(\xp_2^n;\yp_2^n\right)\nn\\
&&\leq
I\left(\xp_1^n;\yp_1^n\right)+I\left(\xp_2^n;\yp_2^n,\Hbf_2\xp_2^n+\np^n\right)\nn\\
&&=h\left(\Hbf_1\xp_1^n+\Hbf_2\xp_2^n+\zp_1^n\right)-h\left(\Hbf_2\xp_2^n+\zp_1^n\right)%
+h\left(\Hbf_2\xp_2^n+\np^n\right)-h\left(\np^n\right)\nn\\
&&\hspace{.15in}+h\left(\Hbf_4\xp_2^n+\zp_2^n\left|\hspace{.05in}\Hbf_2\xp_2^n+\np^n\right.\right)
-h\left(\zp_2^n\left|\hspace{.05in}\np^n\right.\right)\nn\\
&&\stackrel{(a)}=I\left(\xp_1^n,\xp_2^n;\Hbf_1\xp_1^n+\Hbf_2\xp_2^n+\zp_1^n\right)
+h\left(\Hbf_4\xp_2^n+\zp_2^n\left|\hspace{.05in}\Hbf_2\xp_2^n+\np^n\right.\right)
-h\left(\zp_2^n\left|\hspace{.05in}\np^n\right.\right)\nn\\
&&\stackrel{(b)}\leq
I\left(\xp_1^n,\xp_2^n;\Hbf_1\xp_1^n+\Hbf_2\xp_2^n+\zp_1^n\right)
+nh\left(\Hbf_4\bar\xp_2^*+\zp_2\left|\hspace{.05in}\Hbf_2\bar\xp_2^*+\np\right.\right)
-nh\left(\zp_2\left|\hspace{.05in}\np\right.\right)\nn\\
&&\stackrel{(c)}=
I\left(\xp_1^n,\xp_2^n;\Hbf_1\xp_1^n+\Hbf_2\xp_2^n+\zp_1^n\right)
+nh\left(\Hbf_4\bar\xp_2^*+\zp_2\left|\hspace{.05in}\Hbf_2\bar\xp_2^*+\np,\bar\xp_2^*\right.\right)
-nh\left(\zp_2\left|\hspace{.05in}\np\right.\right)\nn\\
&&=I\left(\xp_1^n,\xp_2^n;\Hbf_1\xp_1^n+\Hbf_2\xp_2^n+\zp_1^n\right)\nn\\
&&\leq n\log\left|\Ibf+\Hbf_1\Sbf_1\Hbf_1^\dagger+\Hbf_2\Sbf_2\Hbf_2^\dagger\right|,
\eqa
where
$\zp_i^n=\left[\zp_{i,1}^\dagger,\zp_{i,2}^\dagger,\dots,\zp_{i,n}^\dagger\right]^\dagger$
and $\np^n=\left[\np_1^\dagger,\np_2^\dagger,\dots,\np_n^\dagger\right]^\dagger$, $i=1,2$,
and $\left[\zp_{2,j}^\dagger,\np_j^\dagger\right]^\dagger$, $j=1,\dots,n$, are
i.i.d. as
(\ref{eq:ZICjoint2}).

Equality (a) is from the fact that $\np$ and $\zp_1$ have the same
marginal distribution. Inequality (b) is by Lemma
\ref{lemma:conditionaldirect}, and we let
$\bar\xp_i^*\sim\Cmat\Nmat\left(\0bf,\Sbf_i\right)$, $i=1,2$.
$\bar\xp_1^*$ is independent of $\bar\xp_2^*$ and $\bar\yp_i^*$ is
defined in (\ref{eq:model}) with $\xp_i$ replaced by
$\bar\xp_i^*$. Equality (c) is from (\ref{eq:ZStrongMarkov}) which means
\bqn
\Sbf_2\Hbf_4^\dagger=\Sbf_2\left(\Hbf_2^\dagger\Abf^\dagger+\Bbf^\dagger\right)=
\Sbf_2\Hbf_2^\dagger\Abf^\dagger.
\eqn
By
Lemma \ref{lemma:myMarkovChain},
$\bar\xp^*_2\rightarrow\Hbf_2\bar\xp_2^*+\np\rightarrow\Hbf_4\bar\xp_2^*+\zp_2$
forms a Markov chain.

Therefore, (\ref{eq:ZICstrongRegion}) is an outer bound for the
capacity region. On the other hand, (\ref{eq:ZICstrongRegion}) is
also achievable by requiring user $1$ to decode messages from both
users. Therefore, Theorem \ref{theorem:ZICstrongRegion} is proved.

If $\Hbf_2$ is left-invertible, we can choose
\bqa%
\Abf^\dagger&{}={}&\Hbf_2\left(\Hbf_2^\dagger\Hbf_2\right)^{-1}\left(\Hbf_4^\dagger-\Bbf^\dagger\right),
\eqa
so that (\ref{eq:ZStrongMarkov}) is satisfied. By Lemma
\ref{lemma:leftInvPs}, $\Abf^\dagger\Abf\preceq\Ibf$ is equivalent to (\ref{eq:strongZsimp}). Thus Proposition
\ref{prop:ZICstrongRegion} is proved.

\subsection{Proof of Theorem \ref{theorem:ICstrongRegion} and Proposition \ref{prop:ICstrongRegion}}

Theorem \ref{theorem:ICstrongRegion} can be proved by using
Theorem \ref{theorem:ZICstrongRegion} twice. To prove a converse,
we first remove the
interference link from transmitter $1$ to receiver $2$ and obtain a MIMO ZIC
with $\Hbf_3=\0bf$. The capacity region of the original MIMO IC is
a subset of the capacity region of this MIMO ZIC because we are effectively giving user $1$'s message to
receiver $2$. Theorem \ref{theorem:ZICstrongRegion} gives
the capacity region of this MIMO ZIC with (\ref{eq:strongMarkov2}). Similarly, we remove the
interference link from transmitter $2$ to receiver $1$ and obtain a MIMO ZIC
with $\Hbf_2=\0bf$.
Theorem \ref{theorem:ZICstrongRegion} gives the capacity region of this
MIMO ZIC with (\ref{eq:strongMarkov1}):
\bqa%
\left\{\begin{array}{c}
  R_1\leq\log\left|\Ibf+\Hbf_1\Sbf_1\Hbf_1^\dagger\right| \\
  R_2\leq\log\left|\Ibf+\Hbf_4\Sbf_2\Hbf_4^\dagger\right| \\
  R_1+R_2\leq\log\left|\Ibf+\Hbf_3\Sbf_1\Hbf_3^\dagger+\Hbf_4\Sbf_2\Hbf_4^\dagger\right| \\
\end{array}\right\}.%
\label{eq:Zregion2}%
\eqa

Thus, the capacity region of the original MIMO IC is included in
the intersection of (\ref{eq:ZICstrongRegion}) and
(\ref{eq:Zregion2}) which is (\ref{eq:GICstrongRegion}). On the
other hand (\ref{eq:GICstrongRegion}) is achievable by requiring
both receivers to decode messages from both transmitters, and
therefore (\ref{eq:GICstrongRegion}) is the capacity region.

Proposition \ref{prop:ICstrongRegion} is similarly proved as
Proposition \ref{prop:ZICstrongRegion}.

\subsection{Proof of Theorem \ref{theorem:ZGICsum} and Propositions
\ref{prop:ZICSum} and \ref{prop:generalZICweak}}

 Since $\Abf^\dagger\Abf\preceq\Ibf$, from Lemma \ref{lemma:pd} there exists a Gaussian random
vector $\np$ whose joint distribution with $\zp_2$ is
\bqa%
\left[\begin{array}{c}
  \zp_2 \\
  \np \\
\end{array}\right]\sim\Cmat\Nmat\left(\0bf,\left[\begin{array}{cc}
  \Ibf &\quad \Abf \\
  \Abf^\dagger &\quad \Ibf \\
\end{array}\right]\right).%
\label{eq:ZICjoint}%
\eqa
Moreover, (\ref{eq:ZICweakMarkov}) and (\ref{eq:ZICjoint}) mean
that $\np$ and $\zp_1$ have the same dimension and distribution.

From Fano's inequality, any achievable rates must satisfy
\bqa%
&&n(R_1+R_2)-n\epsilon\nn\\
&&\leq
I\left(\xp_1^n;\yp_1^n\right)+I\left(\xp_2^n;\yp_2^n\right)\nn\\
&&\leq
I\left(\xp_1^n;\yp_1^n\right)+I\left(\xp_2^n;\yp_2^n,\Hbf_2\xp_2^n+\np^n\right)\nn\\
&&=h\left(\Hbf_1\xp_1^n+\Hbf_2\xp_2^n+\zp_1^n\right)-h\left(\Hbf_2\xp_2^n+\zp_1^n\right)%
+h\left(\Hbf_2\xp_2^n+\np^n\right)-h\left(\np^n\right)\nn\\
&&\hspace{.15in}+h\left(\Hbf_4\xp_2^n+\zp_2^n\left|\hspace{.05in}\Hbf_2\xp_2^n+\np^n\right.\right)
-h\left(\zp_2^n\left|\hspace{.05in}\np^n\right.\right)\nn\\
&&\stackrel{(a)}=h\left(\Hbf_1\xp_1^n+\Hbf_2\xp_2^n+\zp_1^n\right)-h\left(\np^n\right)
+h\left(\Hbf_4\xp_2^n+\zp_2^n\left|\hspace{.05in}\Hbf_2\xp_2^n+\np^n\right.\right)
-h\left(\zp_2^n\left|\hspace{.05in}\np^n\right.\right)\nn\\
&&\stackrel{(b)}\leq
nh\left(\Hbf_1\bar\xp_1^*+\Hbf_2\bar\xp_2^*+\zp_1\right)-nh\left(\np\right)
+nh\left(\Hbf_4\bar\xp_2^*+\zp_2\left|\hspace{.05in}\Hbf_2\bar\xp_2^*+\np\right.\right)
-nh\left(\zp_2\left|\hspace{.05in}\np\right.\right)
\label{eq:FanoBound}\\
&&=
nh\left(\Hbf_1\bar\xp_1^*+\Hbf_2\bar\xp_2^*+\zp_1\right)-nh\left(\np\right)
+nh\left(\Hbf_4\bar\xp_2^*+\zp_2\right)+nh\left(\Hbf_2\bar\xp_2^*+\np\left|\hspace{.05in}\Hbf_4\bar\xp_2^*+\zp_2\right.\right)\nn\\%
&&\hspace{.15in} -nh\left(\Hbf_2\bar\xp_2^*+\np\right)
-nh\left(\zp_2\left|\hspace{.05in}\np\right.\right)\nn\\
&&\stackrel{(c)}=
nh\left(\Hbf_1\bar\xp_1^*+\Hbf_2\bar\xp_2^*+\zp_1\right)-nh\left(\np\right)
+nh\left(\Hbf_4\bar\xp_2^*+\zp_2\right)+nh\left(\Hbf_2\bar\xp_2^*+\np\left|\hspace{.05in}\Hbf_4\bar\xp_2^*+\zp_2,\bar\xp_2^*\right.\right)\nn\\%
&&\hspace{.15in} -nh\left(\Hbf_2\bar\xp_2^*+\np\right)
-nh\left(\zp_2\left|\hspace{.05in}\np\right.\right)\nn\\
&&\stackrel{(d)}=
nh\left(\Hbf_1\bar\xp_1^*+\Hbf_2\bar\xp_2^*+\zp_1\right)-nh\left(\np\right)
+nh\left(\Hbf_4\bar\xp_2^*+\zp_2\right)+nh\left(\np\left|\hspace{.05in}\zp_2\right.\right)
-nh\left(\Hbf_2\bar\xp_2^*+\zp_1\right)
-nh\left(\zp_2\left|\hspace{.05in}\np\right.\right)\nn\\
&&=
nh\left(\Hbf_1\bar\xp_1^*+\Hbf_2\bar\xp_2^*+\zp_1\right)-nh\left(\Hbf_2\bar\xp_2^*+\zp_1\right)
+nh\left(\Hbf_4\bar\xp_2^*+\zp_2\right)-nh\left(\zp_2\right)\nn\\
&&=n\log\left|\Ibf+\Hbf_1\Sbf_1\Hbf_1^\dagger\left(\Ibf+\Hbf_2\Sbf_2\Hbf_2^\dagger\right)^{-1}\right|
+n\log\left|\Ibf+\Hbf_4\Sbf_2\Hbf_4^\dagger\right|,%
\label{eq:FanoZIC}
 \eqa
where $\np^n=\left[\np_1^\dagger,\np_2^\dagger,\dots,\np_n^\dagger\right]^\dagger$, and the
$\np_i$ are i.i.d. Gaussian vectors distributed as
(\ref{eq:ZICjoint}).

Equalities (a) and (d) are both from the fact that $\np$ and $\zp_1$
 have the same marginal distribution. Inequality (b) is from
Lemma \ref{lemma:conditionaldirect}, and we let
$\bar\xp_i^*\sim\Cmat\Nmat\left(\0bf,\Sbf_i\right)$, $i=1,2$.
$\bar\xp_1^*$ is independent of $\bar\xp_2^*$ and $\bar\yp_i^*$ is
defined in (\ref{eq:model}) with $\xp_i$ replaced by
$\bar\xp_i^*$.  Equality (c) is from (\ref{eq:ZICweakMarkov}) which means
\bqn
\Sbf_2\Hbf_2^\dagger=\Sbf_2\Hbf_4^\dagger\Abf.
\eqn
 By
Lemma \ref{lemma:myMarkovChain},
$\bar\xp_2^*\rightarrow\Hbf_4\bar\xp_2^*+\zp_2\rightarrow\Hbf_2\bar\xp_2^*+\np$
forms a Markov chain.

Since (\ref{eq:ZICweakSum}) is achievable, the sum-rate capacity
is (\ref{eq:ZICweakSum}) if (\ref{eq:ZICweakMarkov}) holds.
Therefore, Theorem \ref{theorem:ZGICsum} is proved.

When $\Hbf_4$ is left-invertible, we can choose
\bqa%
\Abf&{}={}&\Hbf_4\left(\Hbf_4^\dagger\Hbf_4\right)^{-1}\left(\Hbf_2^\dagger-\Bbf^\dagger\right).
\eqa
Then (\ref{eq:ZICweakMarkov}) is satisfied. By Lemma
\ref{lemma:leftInvPs}, $\Abf^\dagger\Abf\preceq\Ibf$ is equivalent to
(\ref{eq:weakZsimp}), therefore Proposition
\ref{prop:ZICSum} is proved.

Proposition \ref{prop:generalZICweak} is proved in a similar way as Theorem \ref{theorem:ZGICsum}. Let $\widehat\xp_i\sim\Cmat\Nmat\left(\0bf,\widehat\Sbf_i\right)$, $i=1,2$, where
\bqa
\widehat\Sbf_i=\frac{1}{n}\sum_{j=1}^n\Cov\left(\xp_{ij}\right).
\eqa
From Fano's inequality, we have
\bqa%
&&n(R_1+R_2)-n\epsilon\nn\\
&&\leq
I\left(\xp_1^n;\yp_1^n\right)+I\left(\xp_2^n;\yp_2^n,\Hbf_2\xp_2^n+\np^n\right)\nn\\
&&=h\left(\Hbf_1\xp_1^n+\Hbf_2\xp_2^n+\zp_1^n\right)-h\left(\np^n\right)
+h\left(\Hbf_4\xp_2^n+\zp_2^n\left|\hspace{.05in}\Hbf_2\xp_2^n+\np^n\right.\right)
-h\left(\zp_2^n\left|\hspace{.05in}\np^n\right.\right)\nn\\
&&\stackrel{(a)}\leq nh\left(\Hbf_1\widehat\xp_1+\Hbf_2\widehat\xp_2+\zp_1\right)-nh\left(\np\right)
+nh\left(\Hbf_4\widehat\xp_2+\zp_2\left|\hspace{.05in}\Hbf_2\widehat\xp_2+\np\right.\right)
-nh\left(\zp_2\left|\hspace{.05in}\np\right.\right)\nn\\
&&\stackrel{(b)}=nh\left(\Hbf_1\widehat\xp_1+\Hbf_2\widehat\xp_2+\zp_1\right)-nh\left(\np\right)
+nh\left(\Hbf_4\widehat\xp_2+\zp_2\right)
+nh\left(\Hbf_2\widehat\xp_2+\np\left|\hspace{.05in}\Hbf_4\widehat\xp_2+\zp_2,\widehat\xp_2\right.\right)\nn\\%
&&\hspace{.15in} -nh\left(\Hbf_2\widehat\xp_2+\np\right)
-nh\left(\zp_2\left|\hspace{.05in}\np\right.\right)\nn\\
&&=nh\left(\Hbf_1\widehat\xp_1+\Hbf_2\widehat\xp_2+\zp_1\right)-nh\left(\Hbf_2\widehat\xp_2+\np\right)
+nh\left(\Hbf_4\widehat\xp_2+\zp_2\right)-nh\left(\zp_2\right)\nn\\
&&=\log\left|\Ibf+\Hbf_1\widehat\Sbf_1\Hbf_1^\dagger\left(\Ibf+\Hbf_2\widehat\Sbf_2\Hbf_2^\dagger\right)^{-1}\right|
+\log\left|\Ibf+\Hbf_4\widehat\Sbf_2\Hbf_4^\dagger\right|
\label{eq:ZICweakGeneralP}
\eqa
where (a) is from Lemma \ref{lemma:conditionaldirect}; and (b) is from (\ref{eq:ZICgeneral}) which means $\widehat\Sbf_2\Hbf_2^\dagger=\widehat\Sbf_2\Hbf_4^\dagger\Abf$ and thus by Lemma \ref{lemma:myMarkovChain}, $\widehat\xp_2\rightarrow\Hbf_4\widehat\xp_2+\zp_2\rightarrow\Hbf_2\widehat\xp_2+\np$ forms a Markov chain.

\subsection{Proof of Theorem \ref{theorem:NIsum2} and Proposition \ref{prop:NI2}}

Since there exist $\Sigmabf_1$ and $\Sigmabf_2$ which satisfy (\ref{eq:condition2_1}) and (\ref{eq:condition2_2}),  by Lemma \ref{lemma:pd}, there exist two random vectors $\np_1$
and $\np_2$ whose joint distributions with $\zp_1$ and $\zp_2$
are
\bqa%
\left[\begin{array}{c}
  \zp_i \\
  \np_i \\
\end{array}\right]\sim\Cmat\Nmat\left(\0bf,\left[\begin{array}{cc}
  \Ibf &\quad \Abf_i \\
  \Abf_i^\dagger &\quad \Sigmabf_i \\
\end{array}\right]\right),\quad i=1,2.%
\label{eq:NI2ni}%
\eqa
Furthermore, from (\ref{eq:condition2_1}) and (\ref{eq:condition2_2}) we have
\bqa%
\Cov(\np_1)&{}\preceq{}&\Cov\left(\zp_2\left|\hspace{.05in}\np_2\right.\right)\quad\textrm{and}%
\label{eq:NI2Cov1}\\
\Cov(\np_2)&{}\preceq{}&\Cov\left(\zp_1\left|\hspace{.05in}\np_1\right.\right).%
\label{eq:NI2Cov2}%
\eqa

From Fano's inequality, any achievable sum rate $R_1+R_2$ must
satisfy
\bqa %
&&n(R_1+R_2)-n\epsilon\nn\\
&&\leq
I\left(\xp_1^n;\yp_1^n\right)+I\left(\xp_2^n;\yp_2^n\right)\nn\\
&&\leq
I\left(\xp_1^n;\yp_1^n,\Hbf_3\xp_1^n+\np_1^n\right)+I\left(\xp_2^n;\yp_2^n,\Hbf_2\xp_2^n+\np_2^n\right)\nn\\
&&=h\left(\Hbf_3\xp_1^n+\np_1^n\right)-h(\np_1^n)+h\left(\yp_1^n\left|\hspace{.05in}\Hbf_3\xp_1^n+\np_1^n\right.\right)
-h\left(\Hbf_2\xp_2^n+\zp_1^n\left|\hspace{.05in}\np_1^n\right.\right)+h\left(\Hbf_2\xp_2^n+\np_2^n\right)-h(\np_2^n)\nn\\
&&\hspace{.15in}+h\left(\yp_2^n\left|\hspace{.05in}\Hbf_2\xp_2^n+\np_2^n\right.\right)-h\left(\Hbf_3\xp_1^n+\zp_2^n
\left|\hspace{.05in}\np_2^n\right.\right),%
\label{eq:NI2original} %
\eqa
where
$\np_i^n=\left[\np_{i,1}^\dagger,\np_{i,2}^\dagger,\dots,\np_{i,n}^\dagger\right]^\dagger$,
and the $\np_{i,j}$ are i.i.d. Gaussian vectors distributed as
(\ref{eq:NI2ni}). Since $\np_{1,j}$ is independent of $\np_{1,k}$,
and $\zp_{2,j}$ is independent of $\np_{2,k}$, for any $j\neq k$,
from (\ref{eq:NI2Cov1}) we have
\bqa%
\textrm{Cov}\left(\np_{1}^n\right)\preceq\textrm{Cov}\left(\zp_{2}^n\left|\hspace{.05in}\np_{2}^n\right.\right).%
\label{eq:NI2eqncovariance}%
\eqa
By Lemma \ref{lemma:opt} we have
\bqa%
h\left(\Hbf_3\xp_1^n+\np_1^n\right)-h\left(\Hbf_3\xp_1^n+\zp_2^n\left|\hspace{.05in}\np_2^n\right.\right)
\leq nh\left(\Hbf_3\bar\xp_1^*+\np_1\right)-nh\left(\Hbf_3\bar\xp_1^*+\zp_2\left|\hspace{.05in}\np_2\right.\right),%
\label{eq:NI2difference1}%
\eqa
where $\bar\xp_1^*\sim\Cmat\Nmat\left(\0bf,\Sbf_1\right)$. Similarly, we have
\bqa%
h\left(\Hbf_2\xp_2^n+\np_2^n\right)-h\left(\Hbf_2\xp_2^n+\zp_1^n\left|\hspace{.05in}\np_1^n\right.\right)
\leq
nh\left(\Hbf_2\bar\xp_2^*+\np_2\right)-nh\left(\Hbf_2\bar\xp_2^*+\zp_1\left|\hspace{.05in}\np_1\right.\right),
\label{eq:NI2difference2}%
\eqa
where $\bar\xp_2^*\sim\Cmat\Nmat\left(\0bf,\Sbf_2\right)$.

By Lemma \ref{lemma:conditionaldirect} we have
\bqa%
h\left(\yp_1^n\left|\hspace{.05in}\Hbf_3\xp_1^n+\np_1^n\right.\right)
&{}\leq{}& nh\left(\bar\yp_1^*\left|\hspace{.05in}\Hbf_3\bar\xp_{1}^*+\np_1\right.\right)\quad\textrm{and}%
\label{eq:NI2conden1} \\
h\left(\yp_2^n\left|\hspace{.05in}\Hbf_2\xp_2^n+\np_2^n\right.\right)&{}\leq{}&
nh\left(\bar\yp_2^*\left|\hspace{.05in}\Hbf_2\bar\xp_{2}^*+\np_2\right.\right),%
\label{eq:NI2conden2}%
\eqa
where $\bar\yp_i^*$
is defined in (\ref{eq:model}) with $\xp_j$, $j=1,2$, replaced by
$\bar\xp_j^*$.

 On substituting (\ref{eq:NI2difference1})-(\ref{eq:NI2conden2}) into
(\ref{eq:NI2original}) we have
\bqa%
&&R_1+R_2-\epsilon\nn\\
&&\leq
h\left(\Hbf_3\bar\xp_1^*+\np_1\right)-h\left(\np_1\right)+h\left(\bar\yp_1^*\left|\hspace{.05in}\Hbf_3\bar\xp_1^*+\np_1\right.\right)
-h\left(\Hbf_2\bar\xp_2^*+\zp_1\left|\hspace{.05in}\np_1\right.\right)\nn\\
&&\hspace{.15in}+h\left(\Hbf_2\bar\xp_2^*+\np_2\right)-h\left(\np_2\right)+h\left(\bar\yp_2^*\left|\hspace{.05in}\Hbf_2\bar\xp_2^*+\np_2\right.\right)
-h\left(\Hbf_3\bar\xp_1^*+\zp_2\left|\hspace{.05in}\np_2\right.\right)\label{eq:middle}\nn\\
&&=I\left(\bar\xp_1^*;\bar\yp_1^*,\Hbf_3\bar\xp_1^*+\np_1\right)+I\left(\bar\xp_2^*;\bar\yp_2^*,\Hbf_2\bar\xp_2^*+\np_2\right)\nn\\
&&\stackrel{(a)}=I\left(\bar\xp_1^*;\bar\yp_1^*\right)+I\left(\bar\xp_2^*;\bar\yp_2^*\right),\nn\\
&&=\log\left|\Ibf+\Hbf_1\Sbf_1\Hbf_1^\dagger\left(\Ibf+\Hbf_2\Sbf_2\Hbf_2^\dagger\right)^{-1}\right|
+\log\left|\Ibf+\Hbf_4\Sbf_2\Hbf_4^\dagger\left(\Ibf+\Hbf_3\Sbf_1\Hbf_3^\dagger\right)^{-1}\right|,%
\label{eq:NI2Fano}%
\eqa
where (a) is from (\ref{eq:NI2A1}), (\ref{eq:NI2A2}) and Lemma
\ref{lemma:myMarkovChain} since
$\bar\xp_1^*\rightarrow\bar\yp_1^*\rightarrow\Hbf_3\bar\xp_1^*+\np_1$
and
$\bar\xp_2^*\rightarrow\bar\yp_2^*\rightarrow\Hbf_2\bar\xp_2^*+\np_2$
form two Markov chains.

On the other hand (\ref{eq:NI2Fano}) is achievable by treating
interference as noise, and therefore (\ref{eq:NI2Fano}) is the sum-rate
capacity.

Proposition
\ref{prop:NI2} is straightforward from Theorem \ref{theorem:NIsum2}.

\subsection{Proof of Proposition \ref{prop:NIsum} and Proposition \ref{prop:NI1}}

Since matrices $\Abf_1$ and $\Abf_2$ satisfy (\ref{eq:condition}), by Lemma \ref{lemma:solutions} there
exist two Hermitian positive definite matrices $\Sigmabf_1$ and
$\Sigmabf_2$ that satisfy
\bqa%
\Abf_1^\dagger\Abf_1\preceq\Sigmabf_1&{}={}&\Ibf-\Abf_2\Sigmabf_2^{-1}\Abf_2^\dagger\quad\textrm{and}
\label{eq:sigma1}\\
\Abf_2^\dagger\Abf_2\preceq\Sigmabf_2&{}={}&\Ibf-\Abf_1\Sigmabf_1^{-1}\Abf_1^\dagger.%
\label{eq:sigma2}
\eqa
Thus, we see (\ref{eq:condition2_1}) and (\ref{eq:condition2_2}) are satisfied. Since (\ref{eq:NI2A1}) and (\ref{eq:NI2A2}) are satisfied by hypothesis, Proposition \ref{prop:NIsum} follows by Theorem \ref{theorem:NIsum2}.

Proposition \ref{prop:NI1} is straightforward from Proposition \ref{prop:NIsum}.

\subsection{Proof of Theorem \ref{theorem:GICmixed} and Propositions \ref{prop:GICmixed} and \ref{prop:generalMIC}}

The achievability part is straightforward by letting user $2$
first decode the message from user $1$ and then decode its own
message, and by letting user $1$ treat signals from user $2$ as noise.
Then user $1$ and user $2$ have the respective rates
\bqa%
R_1&{}={}&\min\left\{\begin{array}{c}
  \log\left|\Ibf+\Hbf_1\Sbf_1\Hbf_1^\dagger\left(\Ibf+\Hbf_2\Sbf_2\Hbf_2^\dagger\right)^{-1}\right| \\
  \log\left|\Ibf+\Hbf_3\Sbf_1\Hbf_3^\dagger\left(\Ibf+\Hbf_4\Sbf_2\Hbf_4^\dagger\right)^{-1}\right| \\
\end{array}\right\}\quad\textrm{and}%
\nn\\%
R_2&{}={}&\log\left|\Ibf+\Hbf_4\Sbf_2\Hbf_4^\dagger\right|.\nn
\eqa
Therefore, the sum rate (\ref{eq:GICmixed}) is achievable.

To prove the converse, we first let $\Hbf_2=\0bf$. By using
(\ref{eq:mixedMarkov1}) and Theorem
\ref{theorem:ZICstrongRegion}, the sum rate satisfies
\bqa%
R_1+R_2\leq \min\left\{\begin{array}{c}
\log\left|\Ibf+\Hbf_3\Sbf_1\Hbf_3^\dagger+\Hbf_4\Sbf_2\Hbf_4^\dagger\right| \\
  \log\left|\Ibf+\Hbf_1\Sbf_1\Hbf_1^\dagger\right|+\log\left|\Ibf+\Hbf_4\Sbf_2\Hbf_4^\dagger\right| \\
  \end{array}\right\}.%
\label{eq:GICmixed1}%
\eqa
Alternatively, we let $\Hbf_3=\0bf$. By using (\ref{eq:mixedMarkov2})
and Theorem \ref{theorem:ZGICsum}, the
sum rate also satisfies
\bqa%
R_1+R_2\leq
\log\left|\Ibf+\Hbf_1\Sbf_1\Hbf_1^\dagger\left(\Ibf+\Hbf_2\Sbf_2\Hbf_2^\dagger\right)^{-1}\right|
+\log\left|\Ibf+\Hbf_4\Sbf_2\Hbf_4^\dagger\right|.%
\label{eq:GICmixed2}%
\eqa
Combining (\ref{eq:GICmixed1}) and (\ref{eq:GICmixed2}), we have
\bqa%
R_1+R_2\leq\min\left\{\begin{array}{c}
\log\left|\Ibf+\Hbf_3\Sbf_1\Hbf_3^\dagger+\Hbf_4\Sbf_2\Hbf_4^\dagger\right| \\
\log\left|\Ibf+\Hbf_1\Sbf_1\Hbf_1^\dagger\left(\Ibf+\Hbf_2\Sbf_2\Hbf_2^\dagger\right)^{-1}\right|
+\log\left|\Ibf+\Hbf_4\Sbf_2\Hbf_4^\dagger\right|\\
  \log\left|\Ibf+\Hbf_1\Sbf_1\Hbf_1^\dagger\right|+\log\left|\Ibf+\Hbf_4\Sbf_2\Hbf_4^\dagger\right| \\
  \end{array}\right\}.%
\label{eq:GICmixed3}%
\eqa
We complete the proof by pointing out that the last line of
(\ref{eq:GICmixed3}) is redundant because of the second line.

Proposition \ref{prop:GICmixed} is similarly proved by
Propositions \ref{prop:ZICstrongRegion} and \ref{prop:ZICSum}.
Proposition \ref{prop:generalMIC} is similarly proved by
Propositions \ref{prop:generalStrong} and \ref{prop:generalZICweak}.

\subsection{Proof of Theorem \ref{theorem:bound}}
The proof of Theorem \ref{theorem:bound} follows that of Theorem \ref{theorem:ZGICsum}. The bound in problem (\ref{eq:bound}) is derived from (\ref{eq:FanoBound}) by assuming $\bar\xp_i^*\sim\Cmat\Nmat\left(\0bf,\widehat\Sbf_i\right)$, $i=1,2$. Following similar steps as in (\ref{eq:FanoZIC}), one can verify that the sum-rate capacity is (\ref{eq:ZICwekSum}) if (\ref{eq:tightCond}) is satisfied.

\section*{Appendix}
\setcounter{subsection}{0}
\subsection{Proof of Lemma \ref{lemma:generalconcave}}

Let $\xp_{i,S}^*$ be a Gaussian vector with covariance matrix
$\Cov(\xp_{i,S})$. We have
\bqa
\sum_{i=1}^k\lambda_i
h\left(\xp_{i,\Smat}\left|\xp_{i,\Tmat}\right.\right)
&{}\stackrel{(a)}\leq{}&\sum_{i=1}^k\lambda_ih\left(\xp_{i,\Smat}^*\left|\xp_{i,\Tmat}^*\right.\right)\nn\\
&{}={}&\sum_{i=1}^k\lambda_i\left[h\left(\xp_{i,\Smat\cup\Tmat}^*\right)
-h\left(\xp_{i,\Tmat}^*\right)\right]\nn\\
&{}={}&\sum_{i=1}^k\lambda_i\log\left(\frac{\left|\Cov\left(\xp_{i,\Smat\cup\Tmat}^*\right)\right|}
{\left|\Cov\left(\xp_{i,\Tmat}^*\right)\right|}
\cdot\left(\pi e\right)^{\sum_{j\in\Smat}L_j}\right)\nn\\
&{}\stackrel{(b)}\leq{}&\sum_{i=1}^k\log\left(\frac{\left|\Cov\left(\yp_{\Smat\cup\Tmat}\right)\right|}
{\left|\Cov\left(\yp_{\Tmat}\right)\right|}\cdot\left(\pi e\right)^{\sum_{j\in\Smat}L_j}\right)\nn\\
&{}={}&h\left(\yp_{\Smat}\left|\hspace{.05in}\yp_{\Tmat}\right.\right),
\eqa
where inequality (a) is from \cite[Lemma 2]{Thomas:87IT}, and inequality (b) is from \cite[Theorem 17.10.1]{Cover&Thomas:06book}.

\subsection{Proof of Lemma \ref{lemma:conditionaldirect}}

The first inequalities of (\ref{eq:cndEntropy1}) and
(\ref{eq:cndEntropy2}) are straightforward from Lemma
\ref{lemma:generalconcave}. It suffices to prove the second
inequality of (\ref{eq:cndEntropy2}). Since (\ref{eq:cndCov})
holds, we can define two random vectors $\up$ and $\vp$ that are
joint Gaussian, independent of $\widehat\xp^*$ and
$\widehat\yp^*$, and satisfy
\bqa%
\left[\begin{array}{c}
  \bar\xp^* \\
  \bar\yp^* \\
\end{array}\right]=\left[\begin{array}{c}
  \widehat\xp^* \\
  \widehat\yp^* \\
\end{array}\right]+\left[\begin{array}{c}
  \up \\
  \vp \\
\end{array}\right].%
\eqa
Therefore,
\bqa%
h\left(\bar\yp^*\left|\bar\xp^*\right.\right)\geq
h\left(\bar\yp^*\left|\bar\xp^*,\up,\vp\right.\right)
=h\left(\widehat\yp^*\left|\widehat\xp^*\right.\right).%
\eqa

\subsection{ Proof of Lemma \ref{lemma:opt}}
\bqa%
&&h\left(\xp^n+\zp^n\right)-h\left(\xp^n+\zp^n+\tilde\zp^n\right)\nn\\
&&=-I\left(\tilde\zp^n;\xp^n+\zp^n+\tilde\zp^n\right)\nn\\
&&\stackrel{(a)}\leq-I\left(\tilde\zp^n;\xp^{*n}+\zp^n+\tilde\zp^n\right)\nn\\
&&=-h\left(\tilde\zp^n\right)+h\left(\tilde\zp^n\left|\hspace{.05in}\xp^{*n}+\zp^n+\tilde\zp^n\right.\right)\nn\\
&&\stackrel{(b)}\leq
-nh\left(\tilde\zp\right)+nh\left(\tilde\zp\left|\hspace{.05in}\widehat\xp^*+\zp+\tilde\zp\right.\right)%
\nn\\
&&=nh\left(\widehat\xp^*+\zp\right)-nh\left(\widehat\xp^*+\zp+\tilde\zp\right),
\label{eq:imprvWstNoise}
\eqa
where (a) is from \cite[Lemma II.2]{Diggavi&Cover:01IT},  and
$\xp^{*n}$ is a Gaussian vector sequence that has the same
covariance matrix as $\xp^{n}$. Inequality (b) is from Lemma
\ref{lemma:conditionaldirect}. Alternatively, we can use Lemma \ref{lemma:conditionaldirect} to bound (\ref{eq:imprvWstNoise}) as
\bqa
&&nh\left(\widehat\xp^*+\zp\right)-nh\left(\widehat\xp^*+\zp+\tilde\zp\right)\nn\\
&&=-nh\left(\tilde\zp\right)+nh\left(\tilde\zp\left|\hspace{.05in}\widehat\xp^*+\zp+\tilde\zp\right.\right)\nn\\
&&\leq
-nh\left(\tilde\zp\right)+nh\left(\tilde\zp\left|\hspace{.05in}\bar\xp^*+\zp+\tilde\zp\right.\right)\nn\\
&&=
nh\left(\bar\xp^*+\zp\right)-nh\left(\bar\xp^*+\zp+\tilde\zp\right).
\eqa

\subsection{Proof of Lemma \ref{lemma:myMarkovChain}}

Let the eigenvalue decomposition of $\Sbf_u$ be
\bqa%
\Sbf_u=\Qbf\Lambdabf\Qbf^\dagger,%
\label{eq:Covu}%
\eqa
where $\Qbf$ is a unitary matrix and $\Lambdabf$ is a diagonal
matrix with strictly positive diagonal elements. Since
\bqa%
\Cov\left(\xp,\yp\right)\Cov\left(\yp\right)^{-1}
\Cov\left(\yp,\zp\right)=\Cov\left(\xp,\Abf\yp\right)\Cov\left(\Abf\yp\right)^{-1}
\Cov\left(\Abf\yp,\zp\right)
\eqa
for any invertible matrix $\Abf$, we choose $\Abf=\Lambdabf^{-\frac{1}{2}}\Qbf$ and then
$\xp\rightarrow\Hbf\xp+\up\rightarrow\Gbf\xp+\vp$ forms a Markov
chain if and only if
$\xp\rightarrow\tilde\Hbf\xp+\tilde\up\rightarrow\Gbf\xp+\vp$ forms
a Markov chain, where
\bqa%
\tilde\Hbf&{}={}&\Lambdabf^{-\frac{1}{2}}\Qbf\Hbf\quad\textrm{and}%
\label{eq:tildeH}\\
\tilde\up&{}={}&\Lambdabf^{-\frac{1}{2}}\Qbf\up,\nn%
 \eqa
and we have
\bqa%
\Cov\left(\tilde\up\right)&{}={}&\Ibf\quad\textrm{and}\nn\\
\Cov\left(\tilde\up,\vp\right)&{}={}&\Lambdabf^{-\frac{1}{2}}\Qbf\Sbf_{uv}\triangleq\tilde\Sbf_{uv}.
\label{eq:tildeCovuv} %
\eqa

By Lemma \ref{lemma:markov},
$\xp\rightarrow\tilde\Hbf\xp+\tilde\up\rightarrow\Gbf\xp+\vp$ forms
a Markov chain if and only if
\bqa%
\Sbf_x\Gbf^\dagger&{}={}&\Sbf_x\tilde\Hbf^\dagger\left(\Ibf+\tilde\Hbf\Sbf_x\tilde\Hbf^\dagger\right)^{-1}
\left(\tilde\Hbf\Sbf_x\Gbf^\dagger+\tilde\Sbf_{uv}\right)\nn\\
&{}={}&\Sbf_x\tilde\Hbf^\dagger\left(\Ibf+\tilde\Hbf\Sbf_x\tilde\Hbf^\dagger\right)^{-1}
\tilde\Hbf\Sbf_x\Gbf^\dagger+\Sbf_x\tilde\Hbf^\dagger\left(\Ibf+\tilde\Hbf\Sbf_x\tilde\Hbf^\dagger\right)^{-1}\tilde\Sbf_{uv}\nn\\
&{}\stackrel{(a)}={}&\Sbf_x\left[\Ibf-\left(\Ibf+\tilde\Hbf^\dagger\tilde\Hbf\Sbf_x\right)^{-1}\right]\Gbf^\dagger
+\Sbf_x\tilde\Hbf^\dagger\left(\Ibf+\tilde\Hbf\Sbf_x\tilde\Hbf^\dagger\right)^{-1}\tilde\Sbf_{uv}\nn\\
&{}={}&\Sbf_x\Gbf^\dagger-\Sbf_x\left(\Ibf+\tilde\Hbf^\dagger\tilde\Hbf\Sbf_x\right)^{-1}\Gbf^\dagger
+\Sbf_x\tilde\Hbf^\dagger\left(\Ibf+\tilde\Hbf\Sbf_x\tilde\Hbf^\dagger\right)^{-1}\tilde\Sbf_{uv}\nn\\
&{}\stackrel{(b)}={}&
\Sbf_x\Gbf^\dagger-\left(\Ibf+\Sbf_x\tilde\Hbf^\dagger\tilde\Hbf\right)^{-1}
\left(\Sbf_x\Gbf^\dagger-\Sbf_x\tilde\Hbf^\dagger\tilde\Sbf_{uv}\right)\nn\\
&{}\stackrel{(c)}={}&
\Sbf_x\Gbf^\dagger-\left(\Ibf+\Sbf_x\tilde\Hbf^\dagger\tilde\Hbf\right)^{-1}
\left(\Sbf_x\Gbf^\dagger-\Sbf_x\Hbf^\dagger\Sbf_u^{-1}\Sbf_{uv}\right)%
\label{eq:myMarkovChainProof} %
\eqa
where (a) is from the matrix inverse identity \cite[page
151]{Searle:book}
\bqn%
\Abf\left(\Ibf+\Bbf\Abf\right)^{-1}\Bbf
=\Ibf-\left(\Ibf+\Abf\Bbf\right)^{-1}. %
\eqn
Equality (b) is from the matrix inverse identity \cite[page
151]{Searle:book}
\bqn%
\Abf\left(\Ibf+\Bbf\Abf\right)^{-1}=\left(\Ibf+\Abf\Bbf\right)^{-1}\Abf.
\eqn
Equality (c) is from (\ref{eq:Covu}), (\ref{eq:tildeH}) and
(\ref{eq:tildeCovuv}). We complete the proof by pointing out that
(\ref{eq:myMarkovChainProof}) is equivalent to
(\ref{eq:myMarkovChain}).

\subsection{Proof of Lemma \ref{lemma:pd}}

Let $\xp$ be a vector with dimension equal to the number of rows of $\Abf$, and $\yp$ be a vector with dimension equal to the number of columns of $\Abf$. We have $\Bbf\succeq\Abf^\dagger\Abf$ so that $\yp^\dagger\Bbf\yp\geq\yp^\dagger\Abf^\dagger\Abf\yp$ and
\bqa
\left[\begin{array}{c}
        \xp \\
        \yp
      \end{array}
\right]^\dagger\left[\begin{array}{cc}
                       \Ibf &\quad \Abf \\
                       \Abf^\dagger &\quad \Bbf
                     \end{array}
\right]\left[\begin{array}{c}
        \xp \\
        \yp
      \end{array}
\right]&{}={}&\xp^\dagger\xp+\yp^\dagger\Abf^\dagger\xp+\xp^\dagger\Abf\yp+\yp^\dagger\Bbf\yp\nn\\
&{}\geq{}&\xp^\dagger\xp+\yp^\dagger\Abf^\dagger\xp+\xp^\dagger\Abf\yp+\yp^\dagger\Abf^\dagger\Abf\yp\nn\\
&{}={}&\left(\Abf\yp+\xp\right)^\dagger\left(\Abf\yp+\xp\right)\nn\\
&{}\geq{}&0.\nn
\label{eq:pd}
\eqa
Therefore, sufficiency is proved.
On the other hand, if $\left[\begin{array}{cc}
                       \Ibf &\quad \Abf \\
                       \Abf^\dagger &\quad \Bbf
                     \end{array}
\right]\succeq\0bf$, we have
\bqa
\left[\begin{array}{c}
        \xp \\
        \yp
      \end{array}
\right]^\dagger\left[\begin{array}{cc}
                       \Ibf &\quad \Abf \\
                       \Abf^\dagger &\quad \Bbf
                     \end{array}
\right]\left[\begin{array}{c}
        \xp \\
        \yp
      \end{array}
\right]=\xp^\dagger\xp+\yp^\dagger\Abf^\dagger\xp+\xp^\dagger\Abf\yp+\yp^\dagger\Bbf\yp\geq 0.
\label{eq:pdhalf}
\eqa
We choose $\xp=-\Abf\yp$ and substitute it into (\ref{eq:pdhalf}), then we have
\bqa
\yp^\dagger\left(\Bbf-\Abf^\dagger\Abf\right)\yp\geq 0.
\eqa
Therefore, $\Bbf\succeq\Abf^\dagger\Abf$.

If $\Bbf\succ\0bf$, then $\Bbf\succeq\Abf^\dagger\Abf$ is equivalent to
\bqa
0&{}\leq{}&\yp^\dagger\left(\Bbf-\Abf^\dagger\Abf\right)\yp\nn\\
&{}={}&\yp^\dagger\Bbf^{\frac{1}{2}}\left(\Ibf-\Bbf^{-\frac{1}{2}}\Abf^\dagger\Abf\Bbf^{-\frac{1}{2}}\right)
\Bbf^{\frac{1}{2}}\yp\nn\\
&{}={}&\tilde\yp^\dagger\left(\Ibf-\Bbf^{-\frac{1}{2}}\Abf^\dagger\Abf\Bbf^{-\frac{1}{2}}\right)\tilde\yp,
\label{eq:pdhalf2}
\eqa
where
\bqn
\Bbf^{\frac{1}{2}}&{}={}&\Ubf\Lambdabf^{\frac{1}{2}}\Ubf^\dagger\quad\textrm{and}\nn\\
\tilde\yp&{}={}&\Bbf^{\frac{1}{2}}\yp,
\eqn
and
\bqn
\Bbf=\Ubf\Lambdabf\Ubf^\dagger
\eqn
is the eigenvalue decomposition of $\Bbf$ with $\Ubf$ being a unitary matrix and $\Lambdabf$ being a diagonal matrix with strictly positive diagonal elements. Since $\tilde\yp$ can be any vector, (\ref{eq:pdhalf2}) means
\bqn
\Ibf\succeq\Bbf^{-\frac{1}{2}}\Abf^\dagger\Abf\Bbf^{-\frac{1}{2}}.
\eqn
Suppose that the singular value decomposition of $\Bbf^{-\frac{1}{2}}\Abf^\dagger$ is
\bqn
\Bbf^{-\frac{1}{2}}\Abf^\dagger=\Pbf\left[\begin{array}{cc}
                                            \Sigmabf &\quad\0bf \\
                                            \0bf &\quad \0bf
                                          \end{array}
\right]\Qbf^\dagger,
\eqn
where both $\Pbf$ and $\Qbf$ are unitary matrices and $\Sigmabf$ is a diagonal matrix with strictly positive diagonal elements. Then we have
\bqn
\Bbf^{-\frac{1}{2}}\Abf^\dagger\Abf\Bbf^{-\frac{1}{2}}&{}={}&\Pbf\left[\begin{array}{cc}
                                            \Sigmabf &\quad\0bf \\
                                            \0bf &\quad \0bf
                                          \end{array}
\right]\Pbf^\dagger\quad\textrm{and}\\
\Abf\Bbf^{-\frac{1}{2}}\Bbf^{-\frac{1}{2}}\Abf^\dagger&{}={}&\Qbf\left[\begin{array}{cc}
                                            \Sigmabf &\quad\0bf \\
                                            \0bf &\quad \0bf
                                          \end{array}
\right]\Qbf^\dagger.
\eqn
Therefore, $\Ibf\succeq\Bbf^{-\frac{1}{2}}\Abf^\dagger\Abf\Bbf^{-\frac{1}{2}}$ if and only if $\Ibf\succeq\Sigmabf$ which is also the necessary and sufficient condition for
$\Ibf\succeq\Abf\Bbf^{-\frac{1}{2}}\Bbf^{-\frac{1}{2}}\Abf^\dagger=\Abf\Bbf^{-1}\Abf^\dagger$.

\subsection{Proof of Lemma \ref{lemma:leftInvPs}}

Let $\Abf=\Bbf\left(\Bbf^\dagger\Bbf\right)^{-1}\Cbf^\dagger$  and suppose that the
singular value decomposition of $\Bbf$ is
\bqa%
\Bbf=\Ubf\left[\begin{array}{c}
  \Sigmabf \\
  \0bf \\
\end{array}\right]\Vbf^\dagger, %
\label{eq:svdB} %
\eqa
where both $\Ubf$ and $\Vbf$ are unitary matrices, and $\Sigmabf$
is a diagonal matrix with strictly positive diagonal elements.  Suppose
further that
\bqa%
\Ibf&{}\succeq{}&\Abf^\dagger\Abf\nn\\
&{}={}&\Cbf\left(\Bbf^\dagger\Bbf\right)^{-1}\Cbf^\dagger\nn\\
&{}={}&\Cbf\Vbf\Sigmabf^{-2}\Vbf^\dagger\Cbf^\dagger.%
\label{eq:leftInvStart}%
\eqa
Lemma \ref{lemma:pd} implies that $\Xbf^\dagger\Xbf\preceq\Ibf$ if and only if
$\Xbf\Xbf^\dagger\preceq\Ibf$, therefore (\ref{eq:leftInvStart}) is
equivalent to
\bqa%
\Ibf\succeq\Sigmabf^{-1}\Vbf^\dagger\Cbf^\dagger\Cbf\Vbf\Sigmabf^{-1}, %
\eqa
i.e., for any vector $\xp$ we have
\bqa%
0&{}\leq{}&\xp^\dagger\left(\Ibf-\Sigmabf^{-1}\Vbf^\dagger\Cbf^\dagger\Cbf\Vbf\Sigmabf^{-1}\right)\xp\nn\\
&{}={}&\xp^\dagger\Sigmabf^{-1}\Vbf^\dagger\left(\Vbf\Sigmabf^2\Vbf^\dagger-\Cbf^\dagger\Cbf\right)\Vbf\Sigmabf^{-1}\xp\nn\\
&{}={}&\yp^\dagger\left(\Bbf^\dagger\Bbf-\Cbf^\dagger\Cbf\right)\yp,%
\label{eq:leftInvEnd} %
\eqa
where the last line is from (\ref{eq:svdB}), and we define
$\yp=\Vbf\Sigmabf^{-1}\xp$. Since $\xp$ can be any vector and
$\Sigmabf^{-1}\Vbf^\dagger$ is invertible, $\yp$ can also be any vector.
Therefore, (\ref{eq:leftInvEnd}) proves Lemma
\ref{lemma:leftInvPs}.

\subsection{Proof of Lemma \ref{lemma:solutions}}

From (\ref{eq:eqt2}) and the Woodbury matrix identity
\cite{Golub&vanLoanV3:book}:
\bqa%
\left(\Ebf+\Cbf\Bbf\Cbf^\dagger\right)^{-1}=\Ebf^{-1}-\Ebf^{-1}\Cbf\left(\Bbf^{-1}+\Cbf^\dagger\Ebf^{-1}\Cbf\right)^{-1}\Cbf^\dagger\Ebf^{-1},\nn%
\eqa
we have
\bqa%
\Sigmabf_2^{-1}=\Ibf-\Abf_1\left(-\Sigmabf_1+\Abf_1^\dagger\Abf_1\right)^{-1}\Abf_1^\dagger.%
\label{eq:sigma1inv}%
\eqa
Substituting (\ref{eq:sigma1inv}) into (\ref{eq:eqt1}) we have
\bqa%
\Sigmabf_1=\Ibf-\Abf_2\Abf_2^\dagger+\Abf_2\Abf_1\left(\Abf_1^\dagger\Abf_1-\Sigmabf_1\right)^{-1}\Abf_1^\dagger\Abf_2^\dagger.%
\label{eq:eqSigma1}%
\eqa
Define
\bqa%
\Xbf_1&{}={}&\Sigmabf_1-\Abf_1^\dagger\Abf_1,\label{eq:X}\\%
\Mbf_1&{}={}&\Ibf-\Abf_1^\dagger\Abf_1-\Abf_2\Abf_2^\dagger,\label{eq:lemmaM}\\
\Mbf_2&{}={}&\Ibf-\Abf_1\Abf_1^\dagger-\Abf_2^\dagger\Abf_2,\label{eq:lemmaM2}\\
\Wbf_1&{}={}&\Abf_1^\dagger\Abf_2^\dagger\quad\textrm{and}\label{eq:lemmaW1}\\
\Wbf_2&{}={}&\Abf_2^\dagger\Abf_1^\dagger.\label{eq:lemmaW2}%
\eqa
On substituting (\ref{eq:X})-(\ref{eq:lemmaW1}) into
(\ref{eq:eqSigma1}), we have the following matrix equation:
\bqa %
\Xbf_1+\Wbf_1^\dagger\Xbf_1^{-1}\Wbf_1=\Mbf_1.%
\label{eq:mxequation}%
\eqa
Equation (\ref{eq:mxequation}) is a special case of a discrete
algebraic Ricatti equation \cite{Engwerda-etal:93LA&A}. From Lemma
\ref{lemma:pdsolution}, with $\Mbf_1$ Hermitian and positive
definite, (\ref{eq:mxequation}) has a positive definite
solution $\Xbf_1$ (i.e., $\Sigmabf_1\succ\Abf_1^\dagger\Abf_1$) if and
only if
\bqn
\textrm{radius}\left(\Mbf_1^{-\frac{1}{2}}\Wbf_1\Mbf_1^{-\frac{1}{2}}\right)
=\textrm{radius}\left(\Phibf_1\right)\leq\frac{1}{2}.
\eqn

Similarly, applying the Woodbury matrix identity to invert $\Sigmabf_2$ in (\ref{eq:eqt1}) and substituting the result into (\ref{eq:eqt2}), we obtain
\bqa
\Xbf_2+\Wbf_2^\dagger\Xbf_2^{-1}\Wbf_2=\Mbf_2,%
\label{eq:lemmaeqt2}
\eqa
where
\bqn
\Xbf_2=\Sigmabf_2-\Abf_2^\dagger\Abf_2.
\eqn
Matrix equation (\ref{eq:lemmaeqt2}) has a positive definite solution $\Xbf_2$ (i.e., $\Sigmabf_2\succeq\Abf_2^\dagger\Abf_2$) if and only if
\bqn
\textrm{radius}\left(\Mbf_2^{-\frac{1}{2}}\Wbf_2\Mbf_2^{-\frac{1}{2}}\right)=
\textrm{radius}\left(\Phibf_2\right)\leq\frac{1}{2}.%
\eqn

\bibliography{Journal,Conf,Misc,Book}
\bibliographystyle{IEEEbib}
\end{document}